\documentclass[twocolumn,
	prd,
	amssymb,
	preprintnumbers,superscriptaddress,
	nofootinbib]{revtex4-2}

\usepackage[bottom]{footmisc}

\newcommand{\msun}{{\rm M}_\odot}

\usepackage{paralist}
\usepackage{amsmath,amssymb,amsfonts}
\usepackage{verbatim}
\usepackage{graphicx}
\usepackage{rotating}
\usepackage{float}
\usepackage{xcolor, soul}
\usepackage{hyperref}
\usepackage{subfigure}
\usepackage[normalem]{ulem}

\DeclareRobustCommand{\okina}{%
  \raisebox{\dimexpr\fontcharht\font`A-\height}{%
    \scalebox{0.8}{`}%
  }%
}
\pdfoutput=1
\allowdisplaybreaks

\newcommand\aj{AJ}
\renewcommand\apj{{ApJ}}
\newcommand\apjl{{ApJL}}     
\newcommand\apjs{{ApJS}}
\newcommand\aap{{A\&A}}
\newcommand\aaps{{A\&AS}}
\newcommand\mnras{{MNRAS}}
\newcommand\pasp{{PASP}}
\newcommand\ssr{{SSRv}}
\renewcommand\nat{{Nature}}

\begin{document}

\title[Machine Learning the Tip of the Red Giant Branch]{Machine Learning the Tip of the Red Giant Branch}

\author{Mitchell T.~Dennis}
\affiliation{Institute for Astronomy, University of Hawai\okina i at M\=anoa, 2680 Woodlawn Drive,
Honolulu, HI 96822, USA}

\author{Jeremy Sakstein}
\affiliation{Department of Physics \& Astronomy,
University of Hawai\okina i at M\=anoa,
Watanabe Hall, 2505 Correa Road, Honolulu, HI, 96822, USA}
\date{\today}

\label{firstpage}

\begin{abstract}
A method for investigating the sensitivity of the tip of the red giant branch (TRGB) I band magnitude $M_I$ to stellar input physics is presented.~We compute a grid of $\sim$125,000 theoretical stellar models with varying mass, initial helium abundance, and initial metallicity, and train a machine learning emulator to predict $M_I$ as a function of these parameters.~First, our emulator can be used to theoretically predict $M_I$ in a given galaxy using Monte Carlo sampling.~As an example, we predict $M_I = -3.87^{+0.11}_{-0.08}$ in the Large Magellanic Cloud (F20).~Second, our emulator  enables a direct comparison of  theoretical predictions for $M_I$ with empirical calibrations to constrain stellar modeling parameters using Bayesian Markov Chain Monte Carlo methods.~We demonstrate this by using empirical TRGB calibrations to obtain new independent measurements of the metallicity in three galaxies.~We find $\log_{10}(Z)=-2.167^{+0.404}_{-0.492}$ and $\log_{10}(Z)=-2.098^{+0.388}_{-0.528}$ in the Large Magellanic Cloud (F20 and Y19 respectively), $\log_{10}(Z)=-2.146^{+0.400}_{-0.505}$ in NGC 4258, and $\log_{10}(Z)=-2.143^{+0.401}_{-0.508}$ in $\omega$-Centauri.~The LMC and NGC 4258 measurements are consistent with other measurements within $<1\sigma$ errors, and the $\omega$-Centauri measurement are within $<2\sigma$ errors.
\end{abstract}

\maketitle

\section{Introduction}
The tip of the red giant branch (TRGB) is an important tool in modern astronomy.~Its consistency in the I band (Johnsons-Cousins) across different environments enables its use as a standardizable candle \cite{1999IAUS..183...48S, 2019ApJ...882...34F, 2021ApJ...908L...6R}.~Empirically, the TRGB feature in the color-magnitude diagram can be calibrated with errors of order $10\%$ \cite{2019ApJ...886...61Y,  2019ApJ...882...34F, 2020ApJ...891...57F, 2021ApJ...908L...6R, 2022ApJ...932...15A}.~Theoretical predictions from stellar modeling codes are far less precise due to uncertainties affecting the physics of the helium flash coming from environmental factors such as metallicity and the input physics required for stellar modeling, e.g., opacity, mixing length, nuclear reaction rates, and plasma neutrino losses, and large errors in the bolometric corrections needed to compute the I band magnitude, $M_I$ \cite{2017A&A...606A..33S, 2022MNRAS.514.3058S}.~

In the context of the Hubble tension \cite{Abdalla:2022yfr}, different calibrations of the TRGB $M_I$ in the Large Magellanic Cloud (LMC) result in a distance ladder measurement of the Hubble constant, $H_0$, that is either in tension with the value inferred using the cosmic microwave background (CMB) or is  consistent with it \cite{2019ApJ...886...61Y,Soltis:2020gpl,2021arXiv211204510R,2022ApJ...932...15A,2023arXiv230304790A,2019ApJ...882...34F,2020ApJ...891...57F}.~A better theoretical understanding of the TRGB and its associated uncertainties may aid in validating competing empirical calibrations.

One major factor limiting theoretical studies is the run-time needed to simulate stars at the TRGB, which is typically of the order of two hours or more \footnote{Time estimate based on the run-time for the stellar structure code MESA \cite{Paxton2011,Paxton2013,Paxton2015,Paxton2018,Paxton2019,2022arXiv220803651J} using eight cores and 10GB of memory.}~In this work, we present a method to overcome this whereby a large grid ($\sim\!\!125,000$ models) with varying mass, initial helium abundance, and initial metallicity is simulated and used to train a machine learning (ML) emulator.~This emulator evaluates in milliseconds, enabling rapid sampling of parameter space.~While pre-computed interpolated grids for TRGB stars already exist (and also evaluate quickly), a ML emulator (specifically a regression algorithm) offers some advantages over traditional interpolation in higher-dimensional parameter spaces.~Specifically, as the number of dimensions increases, the \textit{curse of dimensionality} dramatically reduces the accuracy of traditional interpolation schemes \cite{2022ScPP...12..187C}.~In contrast, ML regression has been shown to outperform traditional interpolation achieving a much greater accuracy \cite{2022ScPP...12..187C}.~This work is a preliminary feasibility study, and we only explore three parameters in the interest of having a short grid generation time, and to test the ML emulator's accuracy.~Ultimately, we find that the ML is highly accurate, and that more parameters can be varied without degrading this significantly.~As we continue this program and expand into high dimensional parameter space, ML will rapidly become the only viable approach.

We showcase two applications of our emulator.~First, we estimate the theoretical error on the TRGB $M_I$ by performing a Monte Carlo (MC) sampling of the parameter space.~We predict $M_I = -3.87^{+0.11}_{-0.08}$ in the LMC (F20).~Second, we use empirical calibrations of $M_I$ in the LMC (F20 and Y19), $\omega$-Centauri, and NGC 4258 to constrain these objects' metallicity, $Z$, by using Markov Chain Monte Carlo (MCMC) sampling.~Our results are summarized in Table \ref{tbl:results}.~For all of our objects, we find metallicities consistent with other measurements.

A reproduction package \cite{dennis_mitchell_t_2022_7895972} accompanies this work and can be found at the following URL:~\url{https://zenodo.org/record/7895972}.~This includes our entire grid of models, MESA inlists needed to reproduce them, our ML emulators, and our MC and MCMC python scripts used to produce the results presented here.

This paper is organized as follows.~In section \ref{sec:theory} we describe the grid of models used to train the ML emulator.~In section \ref{sec:ML} we describe the ML methods we use to train our emulator.~In section \ref{sec:MC} we use MC sampling to predict the theoretical TRGB $M_I$ in the LMC.~In section \ref{sec:MCMC} we use MCMC sampling to constrain the metallicity in the LMC, NGC 4258, and $\omega$-Centauri.~We discuss our results in section \ref{sec:discussion} and conclude in section \ref{sec:conc}.

\section{Theoretical TRGB Models}
\label{sec:theory}

\subsection{Stellar modeling}
We ran a grid of 124,844 models with varying mass, $M$, initial helium abundance, $Y$, and initial metallicity, $Z$, from the pre-main-sequence to the onset of the helium flash (defined as the time when the power in helium burning exceeds $10^6$ ergs/s) using the stellar structure code -- Modules for Experiments in Stellar Astrophysics MESA version 12778 \cite{Paxton2011,Paxton2013,Paxton2015,Paxton2018,Paxton2019,2022arXiv220803651J}.~The parameters were varied over the ranges $0.7\msun\le M\le2.25\msun$, $0.2\le Y\le  0.3$, and $10^{-5} \le Z\le 0.04$.~The spacing is uniform across parameters, linearly in M (46 points) and Y (46 points) and logarithmically spaced in Z (59 points).~This spacing was chosen to make the training grid as large as possible while keeping its generation time computationally feasible.~It is important that the ML error be subdominant to all other sources to ensure that meaningful measurements of the stellar parameters can be made, but there is no reliable method to determine the number of models needed \textit{a priori}.~The choice to create the largest possible training grid was made to maximize the likelihood that a sufficiently small error would be achieved.~We interrogated the effect of grid size on ML error \textit{a posteriori}, and found that the grid size can be reduced by as much as 50\% before training errors become relevant, implying future work could benefit from either a faster grid generation time or more parameters varying.~See section \ref{sec:resolution} for more details.

Turning to the ranges for the parameters we explore, these were, by design, chosen to be larger than the  anticipated range where a helium flash could occur in order to ensure that we captured the boundary between models that undergo a helium flash and models that shell-flash and/or achieve stable core helium burning \cite{2004sipp.book.....H, 2013sse..book.....K}.~As we will discuss below, it is important to self-consistently determine the entire range of viable parameters to ensure that the MCMC does not become prior-dominated.~In the case of $Y$, we chose a range that does not differ significantly from the primordial helium mass fraction but is broad enough to include possible enhancements from previous generations of stars.~We do not include overshooting in our calculations similar to \cite{2017A&A...606A..33S, 2022MNRAS.514.3058S}, who find that they are a subdominant source of uncertainty compared with the parameters we do vary.~Other stellar modeling parameters corresponding to the choice of input physics were fixed to fiducial values.~In particular, the mixing length $\alpha_{MLT}=1.8$, the nuclear reaction rates were set to the MESA defaults (a combination of NACRE and REACLIB \cite{Angulo:1999zz,2010ApJS..189..240C}), mass loss on the red giant branch (RGB) follows the Reimer's prescription \cite{1975MSRSL...8..369R} with efficiency parameter $\eta=0.1$, and we used the initial elemental abundances reported by  \cite{1998SSRv...85..161G} (GS98).~MESA inlists containing all of our parameter choices can be found in our reproduction package \cite{dennis_mitchell_t_2022_7895972}.
Fixing the input physics was necessary for two reasons.~First, the accuracy of the ML emulator will likely decrease as additional parameters are varied, and it is important that the errors in our analysis are not driven by the errors in the ML.~For this reason, we chose to investigate the effects of parameters where stochastic variation between stars due to mass and environment are expected.~(Said another way, even if other parameters such as mixing length are known perfectly, there will still be variations in $M_I$ due to $M$, $Y$ and $Z$ between different stars.) Wind loss efficiency is also a stochastic variable \cite{1993A&A...272..255J,Jimenez:1996at}, but we did not have sufficient computing power to vary it so, as described in more detail below, we estimated its associated uncertainty using a sparse grid and incorporated this into the MCMC likelihood.~Ultimately, we found that the emulator was highly accurate (see section \ref{sec:ML}); so it is likely that more parameters can be varied without degrading the accuracy.~Our preliminary work is then a proof of concept.~Second, a large number of grid points is necessary to achieve an accurate ML emulator, and each additional parameter must have proportionate representation.~Our grid with three parameters varying consumed one million CPU hours.~Fixing the other parameters enabled the grid generation to be computationally tractable.~We discuss the possibility of varying additional parameters in section \ref{sec:discussion}.~

As noted above, the range of $M$, $Y$, and $Z$ covered by our grid covers the entire parameter space where a core helium flash is expected  \cite{2004sipp.book.....H, 2013sse..book.....K}; however, since the boundary between these objects and other objects is not uniform across parameter space, some of our models do not contribute to the TRGB.~In particular, some  models (low $M$ and $Z$ boundaries) do not reach the TRGB in a time shorter than the age of the universe.~Others burn helium stably in their core before the TRGB (high $M$ and high $Z$ boundaries).~They do not experience a core flash but do experience shell flashes.~Excluding these models from the grid would bias the ML emulator to predict a core flash for parameters that do not give rise to one (within the age of the universe), so we instead train an additional ML emulator that classifies models as either older than the age of the universe, core helium burning, or core helium flashing (see section \ref{sec:ML}).~The former two are assigned zero likelihood in our MCMC analysis (see section \ref{sec:MCMC}) and are excluded from our MC analysis.~We manually exclude these models from the training set for the emulator that predicts $M_I$ and $\Delta M_I$.~A small number of models failed to converge.~These were removed from the grid since they are too few in number to effect the ML.~

We pause briefly to explain our choice to simulate a new grid for this study rather than training an emulator on publicly-available pre-computed grids.~We made this choice for several reasons.~First, we desired precise control over the number of grid points and grid spacing, allowing us to fine-tune and optimize the accuracy of our machine learning algorithm.~This control over the grid construction process ensures that our emulator is highly accurate and reliable.~Second, we aim to cover a large region of parameter space with our grid, rather than focusing on a narrow range of mass and metallicity as is often the case with pre-computed grids.~By encompassing a broader parameter space, our approach prevents the MCMC analysis from being prior-dominated.~Third, by generating our own tracks, we maintain complete control over the entire process, enabling us to accurately track and quantify all sources of error from beginning to end.~This control is crucial for achieving a comprehensive understanding of the uncertainties and their impact on the results.~Finally, future work that uses more advanced machine learning algorithms such as active learning (where the machine learning queries the trainer during training for specific data points, see \cite{2022arXiv220316683A} for an implementation in MESA) would require adjustable grid point selection.~Our grid provides the starting point for such studies.~We note that there may be some science cases where it would be sufficient to train on a pre-computed grid e.g., if only a narrow region of parameter space is required.~In these cases, one could accomplish this using the training algorithm provided in our reproduction package.

\subsection{Bolometric corrections}
\label{sec:newBC}
The theoretical I band magnitude $M_I$ for each grid model and its empirical error $\Delta M_I$, which are what is needed for our MCMC comparison with empirical $M_I$ calibrations, were calculated using the Worthey \& Lee \cite{2011ApJS..193....1W} bolometric correction code, which requires the luminosity $L$, effective temperature $T_{\rm eff}$, surface gravity $\log(g)$, and iron abundance $[{\rm Fe}/{\rm H}]$ as input parameters.~These were calculated using the MESA models --- which are are evolutionary tracks --- by selecting the values corresponding to the model where the Helium flash occurs.

Our reproduction package includes a machine learning emulator (not used in this work\footnote{Our emulator that predicts $M_I$ directly is more accurate.}, see section \ref{sec:ML}) that instead predicts the luminosity $L$, the effective temperature $T_{\rm eff}$, the surface gravity $\log(g)$, and the iron abundance $[{\rm Fe}/{\rm H}]$ \cite{dennis_mitchell_t_2022_7895972}.~These can be used to calculate $M_I$ and $\Delta M_I$ using alternative bolometric corrections e.g., MARCS \cite{2008A&A...486..951G}, or PHOENIX \cite{2008ApJS..178...89D}.~We comment that users interested in using other bolometric corrections could apply them directly to our data set and train a new ML emulator using our codes provided.~This will likely produce more accurate results than applying the bolometric corrections to our ML predictions for the MESA outputs.~

Our method described above is tantamount to assuming that the TRGB $M_I$ is due solely to a single star --- the brightest.~In practice, a population of stars with varying mass and metallicity contribute to $M_I$, which is found using the Sobel edge-detection method applied to the color-magnitude diagram.~Our single star approximation is reasonable because the peak $M_I$ reported by the Sobel edge-detection method can be reasonably assumed to be associated with the brightest object in the system \cite{1993ApJ...417..553L,Ivans:2001qp,Sandquist:2004wf,2013A&A...558A..12V}.~One could go beyond this approximation by instead using our emulator to simulate color-magnitude diagrams by drawing from populations with varying $M$, $Y$, $Z$, and other stellar parameters according to some distribution and applying the same edge-detection method.~The choice of underlying distributions, which for some parameters are unknown, will influence the resultant MCMC bound so a detailed comparison study is warranted.~Furthermore, it is important to characterize and propagate errors in the empirically-determined distributions into the MCMC.~Such a comprehensive study is beyond the scope of this work, but it is our hope to perform such an analysis in future work based on the tools developed here.~Meanwhile, our preliminary proof-of-principle study presented in this paper will determine how feasible this is, and lay the foundation for such a study.

\begin{figure*}
    \centering
    \includegraphics[scale=0.7]{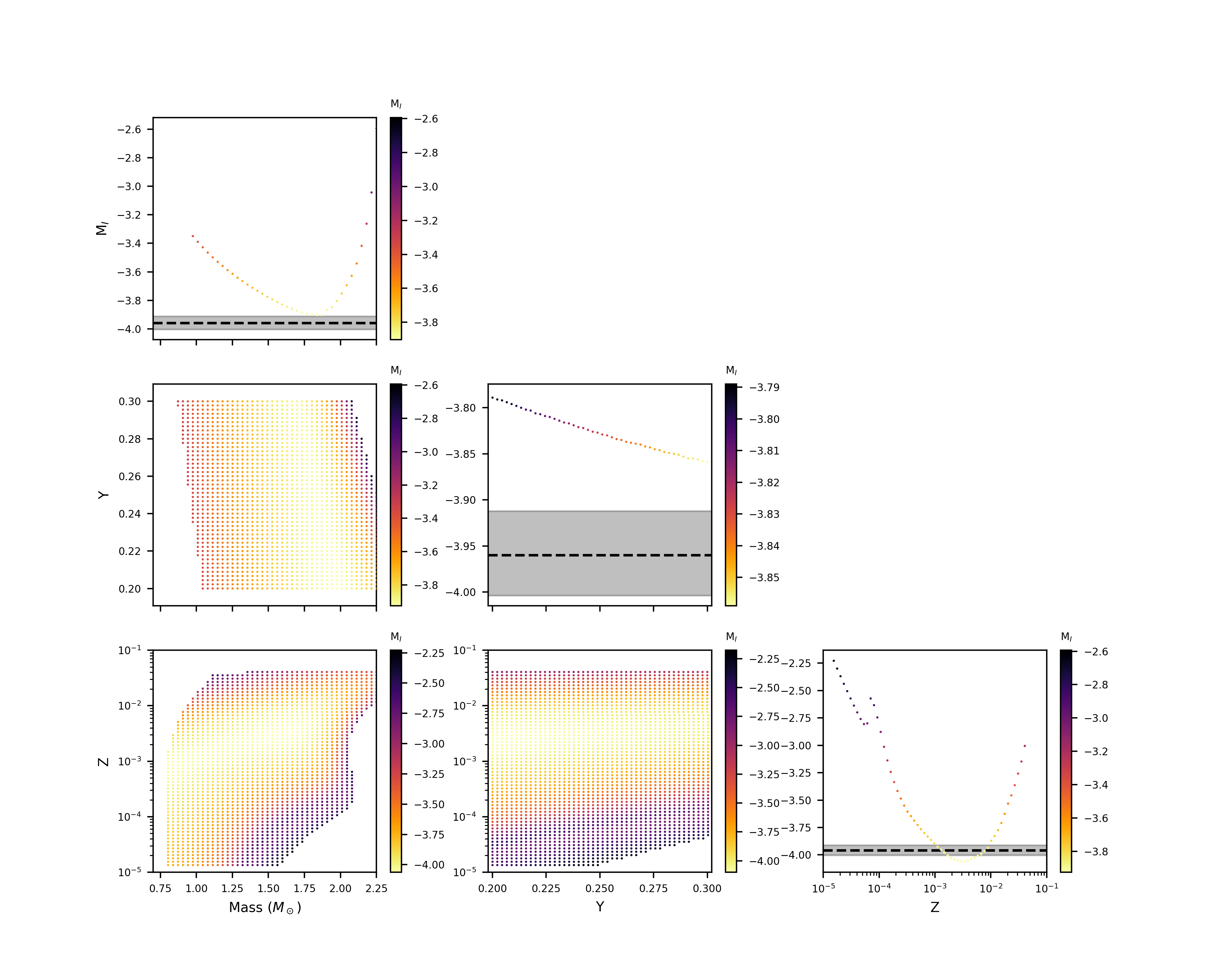}
    \caption{Variation in the TRGB $M_I$ with varying parameters indicated in the corner plot subfigures.~The 2D plots show the $M_I$ values for a fixed $M$, $Y$, or $Z$, depending on the plot axes.~The 1D plots compare the $M_I$ values for two fixed parameters, with the x-axis varying the third parameter.~The x-axis in each column is shared between subplots.~The colorbars for each plot represent the values of $M_I$ where a brighter, more yellow color, indicates a brighter TRGB model.~Color variation in the 1D plots were included for consistency.~The gray horizontal bands in the 1D plots show the \cite{2019ApJ...886...61Y} empirical $M_I$ calibration with $1\sigma$ errors from the LMC.~The parameters were fixed to $M \approx 1.596$, $Y \approx 0.251$, and/or $Z \approx 0.012$.}
    \label{fig:degeneracy}
\end{figure*}

\subsection{Grid of Models}
\label{subsec:gridofmodels}
 The results of our grids (with models that reach the TRGB in a time longer than the age of the universe and models that experience stable core helium burning removed, $\approx 94,000$ models) are exemplified in Figure \ref{fig:degeneracy}.~The figures form a corner plot that show the average value of $M_I$ calibrated empirically for the LMC as given in \cite{2019ApJ...886...61Y} as a function of $M$, $Y$, and/or $Z$.~The x-axes are shared across the columns, and the parameters that are not varied in a given plot are held fixed to $M \approx 1.596$, $Y \approx 0.251$, and/or $Z \approx 0.012$.~The figures illustrate the effects of degeneracies across parameter space for a single 2D slice of our model grid.~A band showing the calibration of $M_I$ in the LMC \cite{2019ApJ...886...61Y} is included in dark grey for reference.~In practice, a single system is unlikely to contain a mixture of stars with such a large range of $M$, $Y$, and $Z$, but it is instructive to examine the trend of $M_I$ across parameter space because our data-driven approach to placing bounds will sample the entire parameter space using MCMC to find the region of maximum likelihood, which is likely to be near the region where the largest number of models intersect the gray band.~The visualization of the trend across the entire parameter space then aids in the physical interpretation of the ultimate  constraints obtained by the MCMC analysis.
 
 Examination of Figure \ref{fig:degeneracy} reveals that there is a weak correlation between initial helium abundance and $M_I$, but the other two parameters show distinct features.~Larger initial masses exhibit a dimmer $M_I$ on average and show a large spread in the possible values.~More stars are rejected as too old or shell flashing at the low edge of mass and the high edge of metallicity, resulting in narrower standard deviations across the respective bins.~As is well-known, at higher masses, the majority of the additional mass becomes part of the stellar envelope, reducing the opacity, which dims the resulting flash \cite{dennis_mitchell_t_2022_7895972}.~Turning to metallicity, there is a peak in brightness around $Z=10^{-2.5}$.~This trend is due to the relation between the bolometric correction in the I band and $T_{\rm eff}$.~Specifically,  the maximum amount of light is emitted into the I band at $T_{\rm eff}\approx 4800$K, which corresponds to the TRGB for stars with $Z\sim 10^{-3}$ \cite{2017A&A...606A..33S,2022MNRAS.514.3058S}.

\subsection{Wind Loss Uncertainties}
\label{sec:wind_uncertainties}

As discussed above, wind loss is also a stochastic variable affecting $M_{I,\,{\rm TRGB}}$ but we were unable to include it in our modeling grid due to computational resource constraints.~To estimate its uncertainty, we simulated a sparse grid of models which fixed the same input physics as our previous model grid, but with varying $M$,  $Z$, and mass loss (Reimers $\eta$ parameter).~To preserve computational resources, $M$ and $Z$ were varied over the ranges $0.78 M_\odot - 1.8 M_\odot$, and $Z = 0.00031 - 0.0222$ respectively.~Because of the low variation in $Y$, we held it fixed at $Y=0.245$.~These correspond to the ranges we varied in our initial parameter grid.~Wind loss was varied over the range $0\le\eta\le0.2$.

\begin{figure}
    \centering
    \includegraphics[scale=0.5]{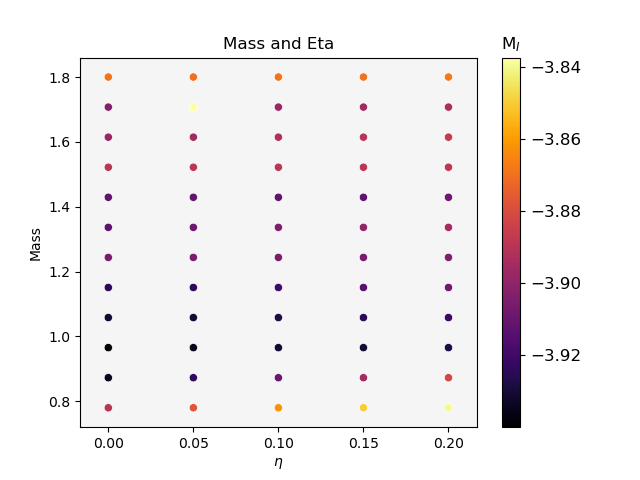}
    \includegraphics[scale=0.5]{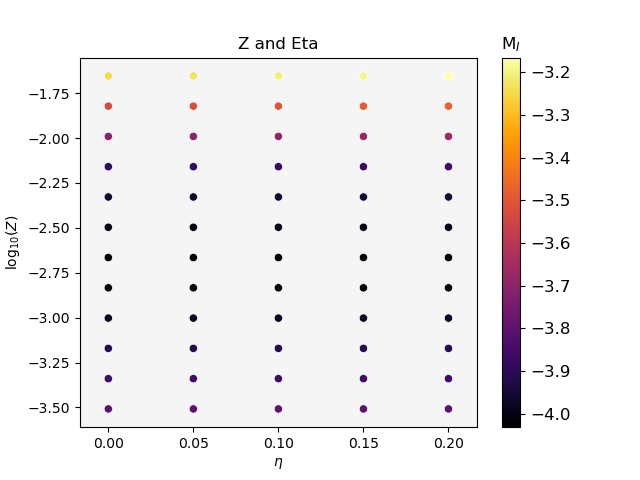}
    \caption{Variation in $M_I$ as a function of mass and $\eta$ (top plot) and $Z$ and $\eta$ (bottom) that varied initial mass,  metallicity, and wind loss over $0.78 M_\odot - 1.8 M_\odot$, and $Z = 0.00031 - 0.0222$.~Inspection of the Z plot indicates an increase in $\Delta\textrm{M}_I$ as a function of $\eta$ (the spread between the largest and smallest values gets wider).~There is also a weak trend in mass.}
    \label{fig:derivative_plots}
\end{figure}

We calculated the spread in $ M_{I}$ due to the variation in $\eta$, $\Delta M_{I,\eta}$, by taking the median of the difference between the $M_I$s for a fixed $\eta$ and $Z$ (or mass).~This gives the median $\Delta M_{I,\eta}$ as a function of $\eta$ and $Z$ (or $\eta$ and $M$).~We conservatively used the maximum of these medians as the value of $\Delta M_{I,\eta}$.~To find the associated error for $\Delta M_{I,\eta}$ that is used in the MCMC, we assumed a uniform distribution and calculated the error with $\sigma_\eta = \sqrt{\frac{\Delta M_{I,\eta}}{12}}$.~We took the maximum value of the error, which was $\Delta M_{I,\eta} = 0.12$, and subsequently added it in quadrature to the error in $M_I$.~These errors are subdominant to those coming from variations in  M, Z, and to the uncertainties in the Bolometric corrections.~The variation in the $M_I$ values are shown in figure ~\ref{fig:derivative_plots}.

\section{Machine Learning Emulator}
\subsection{DNN Components and Training}
\label{sec:ML}

\begin{figure*}
    \centering
    \includegraphics[scale=0.5]{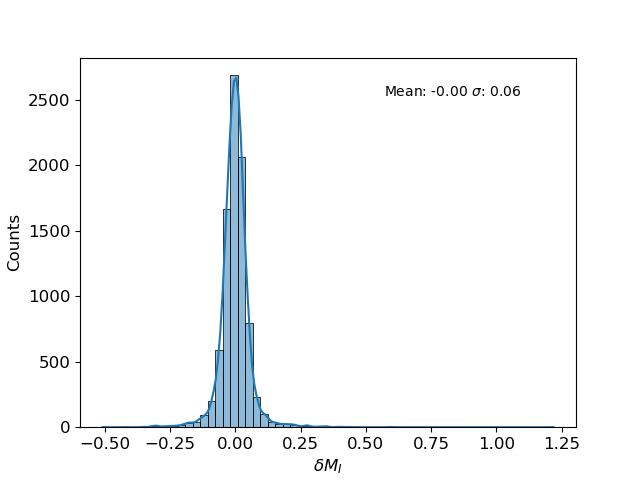}
    \includegraphics[scale=0.5]{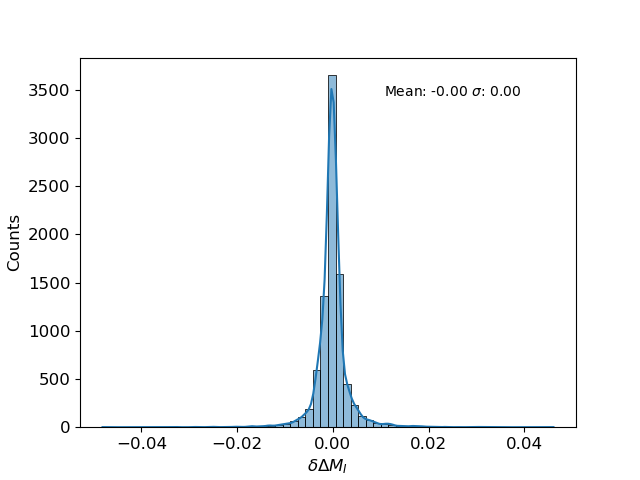}
    \caption{Error distributions for $M_I$, $\Delta M_I$ from the ML regression.~These distributions were found by subtracting the ML predictions from the true values found by applying the bolometric correction code to each grid point that reached the TRGB in a time shorter than the age of the universe.~These distributions represent the results from the testing data, i.e., data that the network was not trained on.}
    \label{fig:mError}
\end{figure*}

We trained a ML emulator on the grid of models described in the previous section.~~We used a deep neural network (DNN) with the tensorflow \cite{tensorflow2015-whitepaper} and keras packages \cite{chollet2015keras} to build the emulator because of advantages DNNs hold over other ML methods in high dimensional parameter spaces which we discuss further in section \ref{subsec:comparison}.~We implemented the network using the ADAM \cite{2014arXiv1412.6980K} optimizer and hand tuned the hyperparameters of the network to optimize the training results.~We remark that other methods, e.g., explored in \cite{2019arXiv191206059L} may be necessary for higher dimensional data sets.~The ML emulator was built in two components, a classifier that evaluates whether or not a set of input parameters $\{M,Y,Z\}$ will yield a star on the TRGB (or whether they will exhibit stable core helium burning, or reach TRGB in a time longer than the age of the universe), and a regression algorithm that predicts $M_I$ and the empirical error in the I band for given $\{M,Y,Z\}$.~Previous work that applied machine learning to other stellar objects \cite{2017EPJWC.16005003B, 2016ApJ...830...31B} used a random forest to emulate input parameters from stellar observables (the opposite of our procedure).~Our procedure differs from theirs in that our emulator contains a classification component.~Further, our ML algorithm is called within the MCMC rather than as a separate step in the data analysis pipeline.~

We also trained a separate ML regression algorithm that takes $\{M,Y,Z\}$ as inputs, but instead outputs the raw MESA outputs, luminosity $L$, effective temperature $T_{\rm eff}$, surface gravity $\log(g)$, and iron abundance $[{\rm Fe}/{\rm H}]$.~This emulator had a longer evaluation time and was less accurate than the emulator above so was not used in this work.~We provide this as part of our reproduction package for readers interested in using different bolometric corrections \cite{dennis_mitchell_t_2022_7895972}.
Both algorithms used an 80\%/10\%/10\% split between training, validation, and testing data, but the training data set for the classifier was re-sampled using SMOTE \cite{2011arXiv1106.1813C} due to the large imbalance between the three classes which can have a negative effect on the network \cite{2019SPIE11198E..13S}.~SMOTE creates synthetic data by taking convex combinations of groups of data points in the minority class, randomly sampling within the convex combinations until the all the classes have an equal number of data points.~This yields a balanced data set which is better for the ML training because an unbalanced dataset can bias the network towards labeling more objects as the more populous class.

\subsection{Emulator Errors}

\label{sec:emulator_errors}

The classification algorithm has an accuracy of 99.7\% and a categorical crossentropy loss of 0.01 across the three categories.~The regression algorithm used predicts $M_I$ and $\Delta M_I$ with a mean squared error loss of $3.58 \times 10^{-5}$ for input data and corresponding output values normalized between 0 and 1.~Histograms of the residuals from the testing data set for $M_I$ and $\Delta M_I$ are shown in Figure \ref{fig:mError}.~The errors in our ML algorithm are highly subdominant to those coming from the empirical bolometric corrections, which have an average error of order $0.1$ mag.

\begin{figure*}
    \centering
    \includegraphics[scale=0.7]{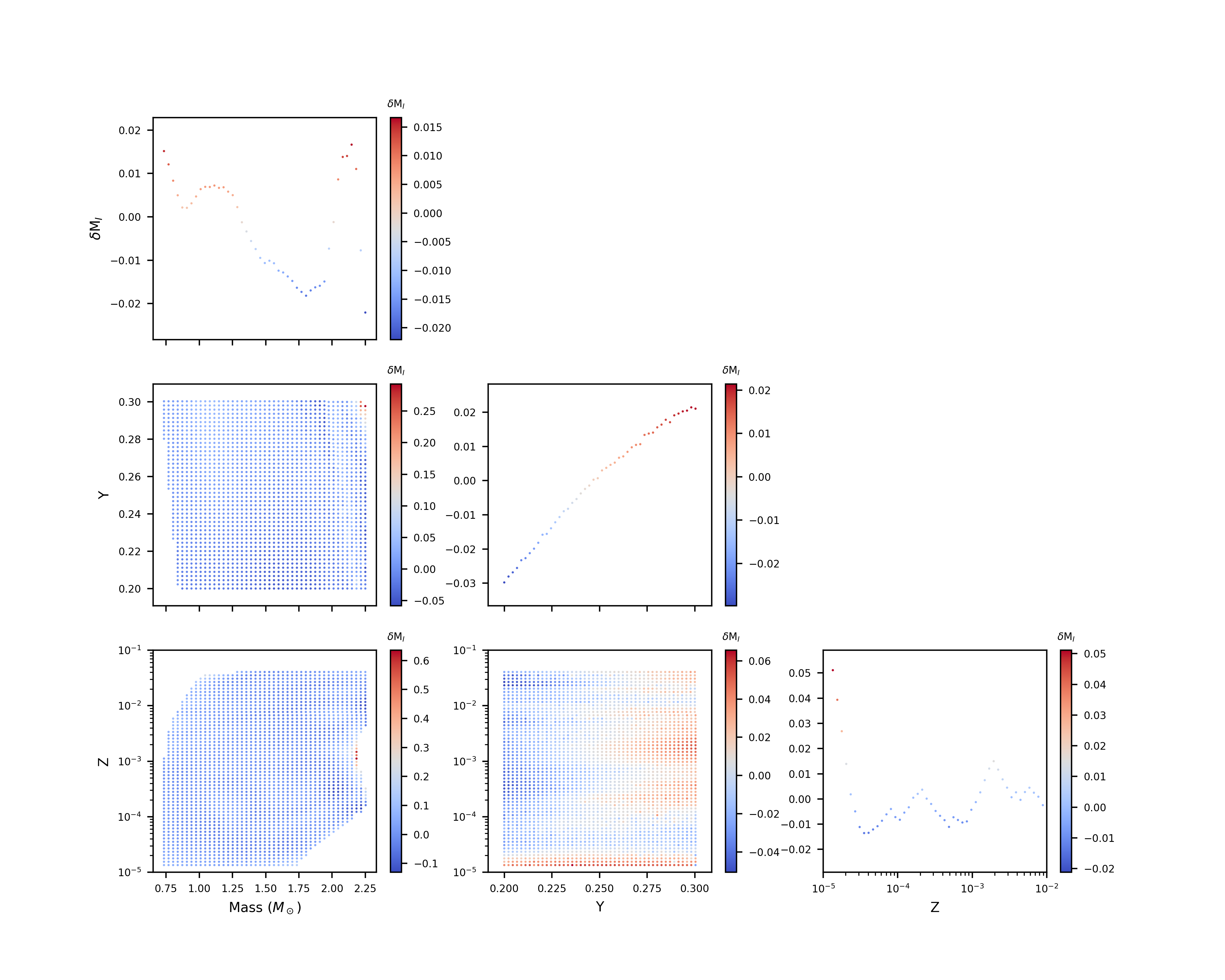}
    \caption{The residual plots for the MESA emulator and the training data.~For the 2D plots, the residual was averaged over the third axis.~For the 1D plots, the residual was averaged over two axes.~In this case we used the median to calculate the average to avoid skew caused by outliers.~The colorbars represent the values of the residual, and we comment that the most extreme values of error are almost exclusively confined to areas the MCMC samplers do not favour.~These errors are subdominant to the errors given by the bolometric corrections (order 0.1).~This finding is similar to the result found in Figure \ref{fig:mError} which displays the distributions of the residuals for the outputs of our emulator and the testing data.}
    \label{fig:cornerErrors}
\end{figure*}

In addition to these distributions, we present the residuals from the training data and the results from our emulator.~These are presented in \ref{fig:cornerErrors}, a corner plot that plots the residual as a function of two parameters, averaged over a third parameter.~For this average we again used the median to prevent outlier bias.~The individual plots are dominated by markers that represent near zero error values.~Further, when figure \ref{fig:cornerErrors} is compared to figure \ref{fig:degeneracy}, we can see the parameters with the brightest stellar values, i.e., parameters where the MCMC is more likely to sample, 
are areas where the error is correspondingly low.~We caution the reader to note that the color bar values (e.g.~the errors) change for each member plot in figure \ref{fig:cornerErrors}, and so a color bar for each subplot has been provided.

\subsection{Comparison to Other Machine Learning Methods}
\label{subsec:comparison}
A deep neural network is not the only viable method for creating an emulator for MESA.~Other models such as decision trees, random forests, gradient boosting, or k-nearest neighbors, among others, can also be used to emulate MESA.~We tested several alternative algorithms and compared their mean-squared errors  (MSEs).~A summary of the methods we tested, and their corresponding MSE accuracy are given in Table \ref{tbl:accuracyComparison}.

\begin{table}[h]
    \centering
    \begin{tabular}{|c|c|}
         \hline Method & MSE Score \\
         \hline Linear Regression & $5.3 \times 10^{-3}$  \\
         \hline Support Vector Machine & $4.6 \times 10^{-3}$ \\
         \hline SVM with Stochastic Gradient Descent & $5.3 \times 10^{-3}$ \\
         \hline K-Nearest Neighbors & $6.00\times 10^{-5}$ \\
         \hline Decision Tree & $1.02 \times 10^{-5}$ \\
         \hline Gradient Boosted Decision Tree & $2.2 \times 10^{-4}$ \\
         \hline Random Forest & $6.0 \times 10^{-6}$ \\
         \hline Extreme Gradient Boosting & $2.6 \times 10^{-5}$ \\
         \hline Deep Neural Network [this work] & $3.6 \times 10^{-5}$ \\
         \hline
    \end{tabular}
    \caption{The results of fitting training data to multiple different machine learning methods and scoring them with the testing data.}
    \label{tbl:accuracyComparison}
\end{table}

The results from Table \ref{tbl:accuracyComparison} indicate that the highest performing method for emulating MESA is the random forest, followed by its constituent member method, a decision tree.~These are followed closely by gradient boosted decision trees (XGBoost), our deep neural network, and k-nearest neighbors, respectively.~Of these top performers, random forest stands out as performing an order of magnitude better than the other methods.~This result is unsurprising; random forests have consistently outperformed neural networks on tabular data \cite{2022arXiv220708815G}, even with less data.~However, random forests and other methods often reach a plateau once a certain amount of data is reached \cite{ng.machinelearningyearning}, whereas deep neural network continue to improve.~Beyond this, algorithms such as k-nearest neighbors and support vector machines (SVM)s are both computationally intensive, particularly for large datasets, and suffer from the \textit{curse of dimensionality}.~Lastly, among the models we tested only deep neural networks are differentiable and have a comparably low mean squared error to random forests.~Differentiability is necessary to utilize advanced MCMC methods such as Hamiltonian MCMC (HMCMC) which are more efficient, particularly in high dimensional parameter spaces \cite{2014arXiv1410.5110B, 2017arXiv170102434B, 2020arXiv200700180P, 2022EP&S...74...86Y}.~Looking forward, as we expand this program to varying more parameters, which will necessitate the use of HMCMC, the benefits of DNNs outweigh those of the other algorithms described above, hence our choice to use this algorithm.

\subsection{Training Grid Resolution}
\label{sec:resolution}

As mentioned above, there is no reliable way to determine the number of grid points needed to accurately train a machine learning algorithm \textit{a priori}, so we created the largest training set possible.~To facilitate future studies, we performed resolution tests on our training data for our deep neural networks  to identify the minimum number of models needed for efficient training, defined as the smallest number of grid points needed for the machine learning error to be subdominant to all other sources.

\begin{figure*}
    \centering
    \includegraphics[scale=0.5]{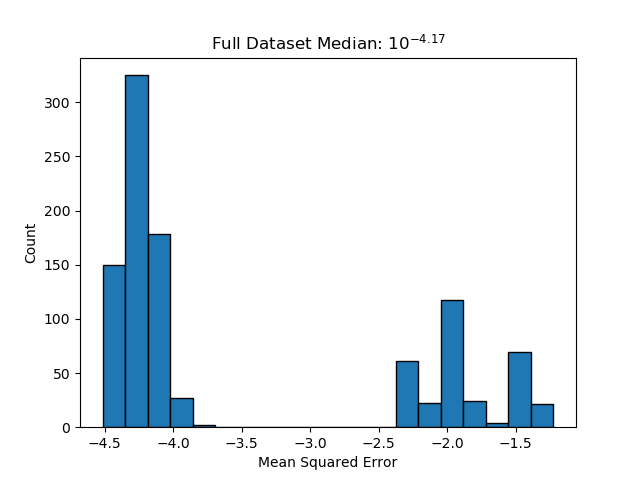}
    \includegraphics[scale=0.5]{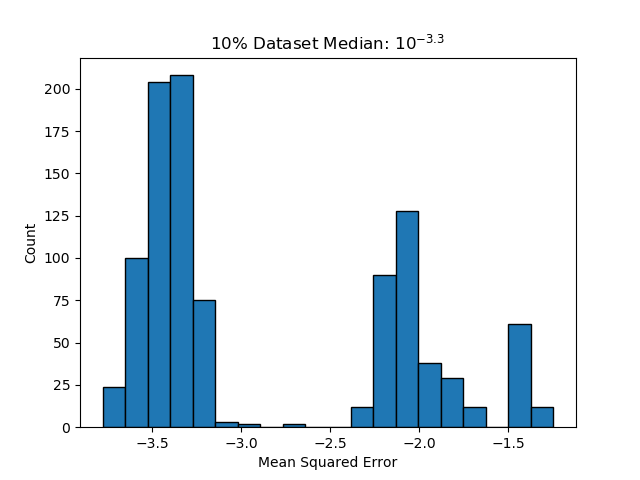}
    \includegraphics[scale=0.5]{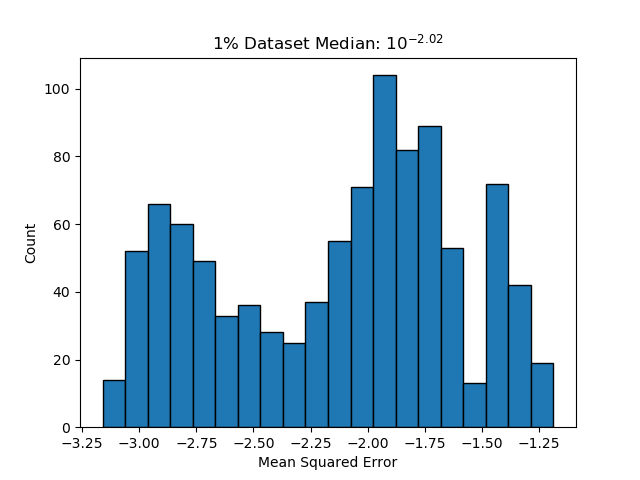}
    \includegraphics[scale=0.5]{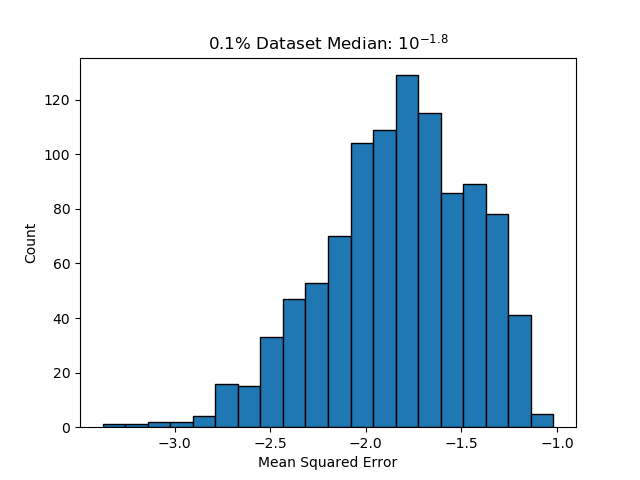}
    \caption{The distribution of the mean squared errors for the 1,000 individual models trained using different sized training sets.~The subset size and median are given in the title of each subplot.~The median was used because of the skewed nature of these plots, and a logarithmic scaling is used to better illustrate the clustering at specific orders of magnitude.~Three of the subfigures show a bimodal distribution.~This behavior is not unexpected because each of the 1000 models is trained with a different random selection of the training data, because each model is initialized with random weights, and because we utilize a learning rate monitor to lower the learning rate under certain circumstances.~Each of these can cause the bimodal behavior.~This is explored in more detail in \S \ref{sec:resolution}.}
    \label{fig:errorTests}
\end{figure*}

We took sequentially smaller subsets of the data, and trained neural networks on these smaller sets.~Due to the randomness in deep neural networks, it is possible to get a variation in the accuracy results.~To account for this, we trained 1,000 networks on 1,000 randomly sampled variations of the data for each fractional subsample.~We tested subsets of 100\% (i.e., no change), 10\%, 1\%, and 0.1\%.~The results of these tests are included as histograms in Figure \ref{fig:errorTests}.~These figures show that for all samples sizes, there are always some models that perform poorly $\approx 10^{-2}$.~However, as the size of the dataset shrinks, the fraction of models that perform poorly increases until at a dataset size of $0.1-1\%$, these poorly performing models dominate.~

Three of the subpanels in Figure~\ref{fig:errorTests} display bimodal error distributions.~These arise from inherent randomness in the deep neural network (DNN) construction and training process, as well as from specific strategies we employ to enhance training.~Each of the 1000 networks is initialized with random weights and trained on a randomly sampled subset of the data.~Some combinations of initial weights and training data are less effective than others, leading to a population of networks with systematically higher errors.

In addition, we employ a learning rate scheduler that reduces the learning rate when validation loss plateaus.~This technique typically improves convergence by allowing the network to explore the loss surface at finer resolution \cite{2019arXiv191011605M}.~However, if the learning rate is reduced too early, the network can become trapped in a suboptimal region of parameter space, resulting in reduced performance.~To balance performance and training efficiency, we optimized the learning rate schedule and number of epochs for typical use cases.~Consequently, a small fraction of networks inevitably underperformed.~In practice, training a modest number (5-10) networks is sufficient to obtain at least one with acceptable accuracy.~We recommend that users train an ensemble and select the best-performing model(s) based on validation error.

\begin{figure*}
    \centering
    \includegraphics[scale=0.5]{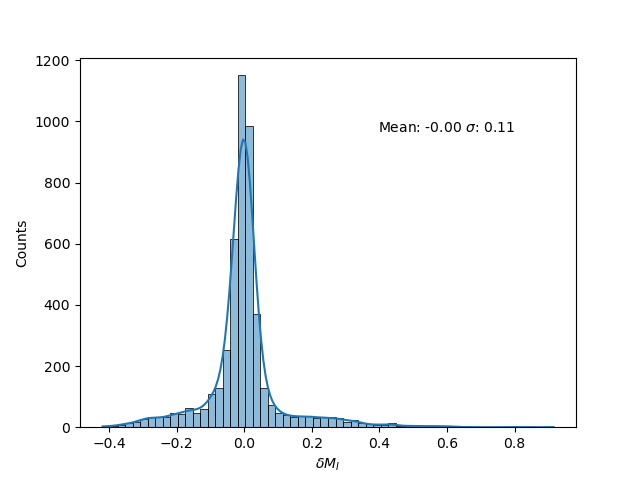}
    \includegraphics[scale=0.5]{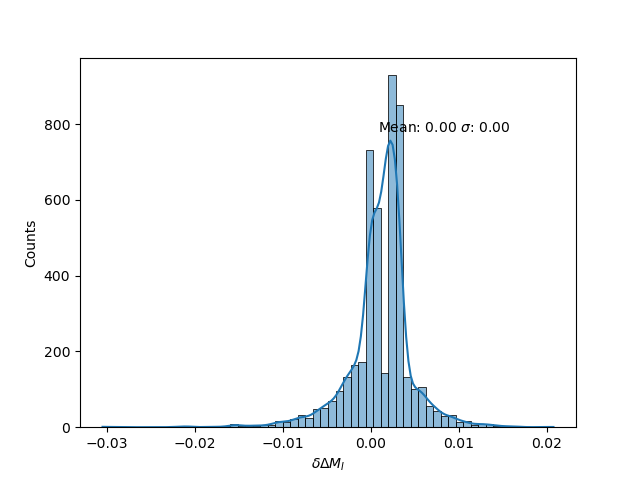}
    \includegraphics[scale=0.5]{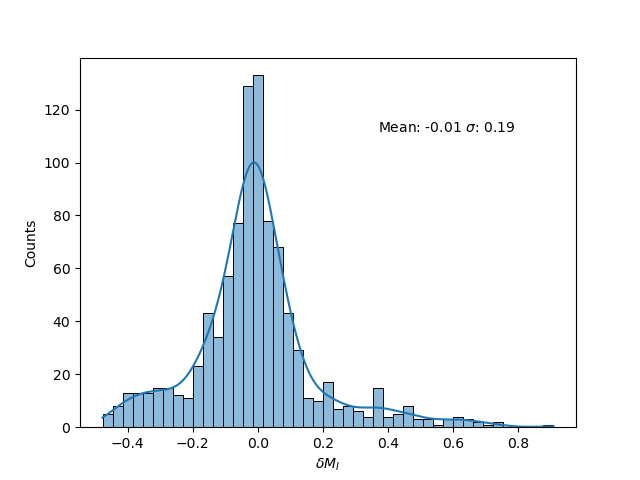}
    \includegraphics[scale=0.5]{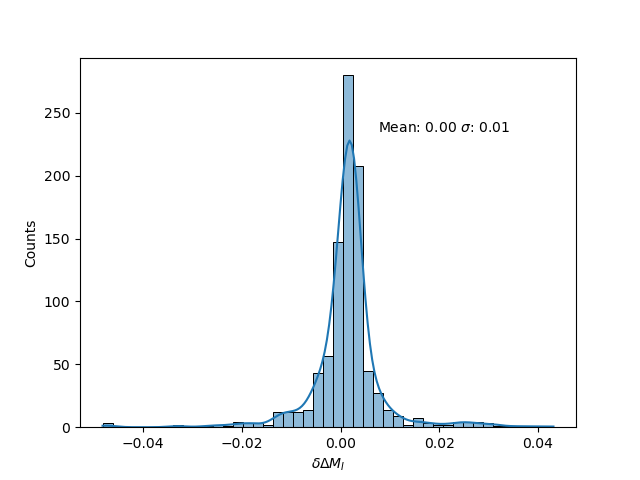}
    \includegraphics[scale=0.5]{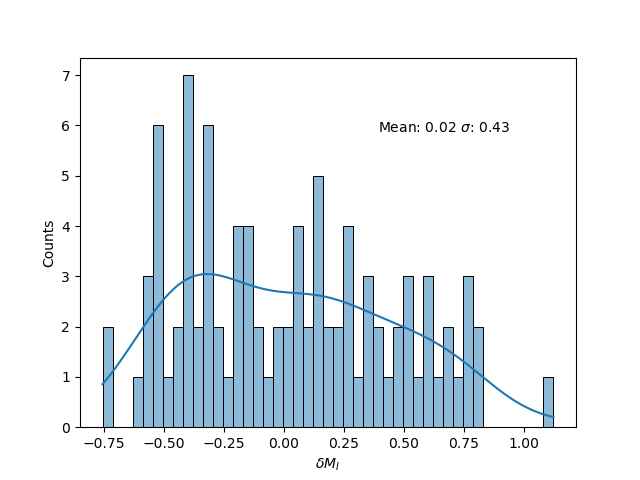}
    \includegraphics[scale=0.5]{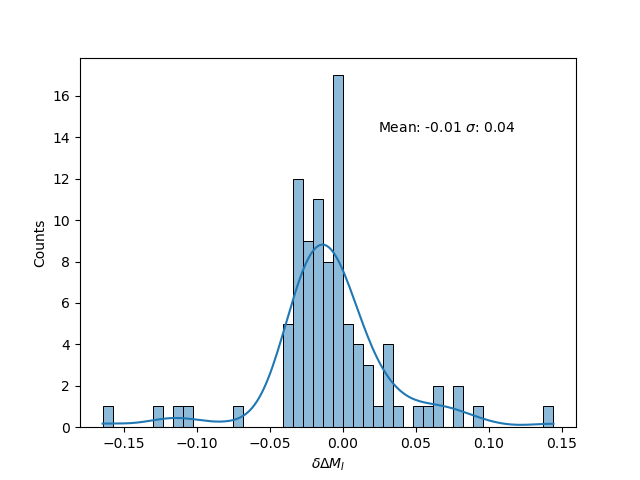}
    \caption{The same as Figure \ref{fig:mError}, but for 50\% (first row), 10\% (second row) and 1\% (third row) of the data.}
    \label{fig:resolution_error}
\end{figure*}

The subplots in Figure \ref{fig:errorTests} are for normalized outputs (i.e.~the outputs are between 0-1) so we also produced plots similar to those in figure \ref{fig:mError} for a single typical networks at 50\%, 10\%, and 1\% of the data shown in figure ~\ref{fig:resolution_error}.~These show that the accuracy begins to degrade significantly when roughly 50\% of the data are removed.~Specifically, the number of models needed to achieve a similar accuracy to ours is roughly same order of magnitude as our full dataset of 90,000 models (around 50,000).

\section{Results}

\begin{figure*}
    \centering
    \includegraphics[scale=0.45]{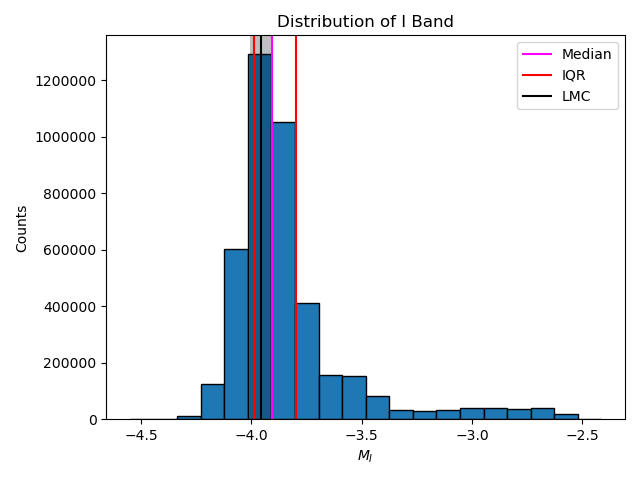}
    \includegraphics[scale=0.45]{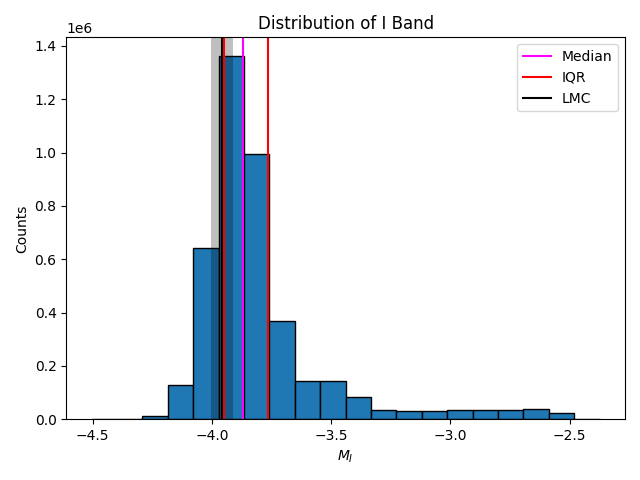}
    \caption{The distribution of $M_I$ after Monte Carlo sampling was performed on the parameter space $\{M,Y,Z\}$ and passed through the ML network (including error accounting).~The gray band shows the zero point $M_I$ in the LMC reported by \cite{2019ApJ...886...61Y} (with 1$\sigma$ errors) and the magenta lines (IQR) indicate the interquartile range.~The prior distributions were taken to be uniform on all parameters (left panel) and uniform on $M$ and $Y$ with a Gaussian prior on the metallicity $Z=0.0091\pm0.0007$ (right panel).}
    \label{fig:iBandDistSM}
\end{figure*}

\subsection{Theoretical I Band Calibration}
\label{sec:MC}

In this section we estimate the theoretical error in $M_I$, focusing on the LMC as an example, due to variations in $M$, $Y$, and $Z$ by performing Monte Carlo sampling.~Specifically, we randomly draw values of $M$, $Y$, and $Z$ from an assumed underlying distribution and use our ML emulator to predict $M_I$.~Repeating this process a number of times builds up a distribution of $M_I$ whose statistics (mean and interquartile range) we then interpret.~We performed 5,000,000 random draws assuming each parameter is uniformly distributed across our parameter range i.e., no prior knowledge of the input parameters was assumed except for the ML classifier which excludes models that do not contribute the TRGB.~For each draw, the empirical error in the bolometric correction was accounted for by drawing from a Gaussian distribution with zero mean and standard deviation $\Delta M_I$ predicted by our emulator for each point and adding it to the predicted $M_I$.~The errors in the ML predictions were accounted for by drawing from the distributions shown in Figure \ref{fig:mError} and adding the result to the corresponding ML prediction.~The error in the ML is highly subdominant to the empirical bolometric correction errors.~The resulting distribution of $M_I$ is shown in the left panel of Figure \ref{fig:iBandDistSM}.~We predict $M_I = -3.91^{+0.11}_{-0.08}$ (median and interquartile range\footnote{We report the median and interquartile range rather than the mean and standard deviation due to the highly non-Gaussian distribution.~The mean and standard deviation are strongly biased by the long tail at high $M_I$}).~This prediction is consistent with the value of $M_I$ reported by \cite{2019ApJ...886...61Y}.

As discussed in Section~\ref{subsec:gridofmodels}, our method above assumes that $M_I$ is determined by the brightest star on the TRGB.~While the Sobel edge-detection method used to calibrate $M_I$ operates on the full color-magnitude diagram and includes contributions from stars near the TRGB edge, the observable is typically dominated by the most luminous objects \cite{1993ApJ...417..553L,Ivans:2001qp,Sandquist:2004wf,2013A&A...558A..12V}.~A more complete treatment would involve simulating full stellar populations spanning distributions in $M$, $Y$, and $Z$, generating synthetic CMDs, and applying the same edge-detection procedure to identify the TRGB.~This is computationally expensive with traditional stellar evolution codes, but our machine learning emulator enables rapid generation of such synthetic populations.~This opens the door to more detailed modeling efforts, including marginalization over population-level properties and full propagation of uncertainties.~We defer such a study to future work, where additional stellar parameters can be varied and emulated, but the present analysis lays the groundwork for that next step.

Our analysis above made the most parsimonious choice for the underlying distributions for $M$, $Y$, and $Z$, namely that they are distributed uniformly.~This has the benefit that the resultant theoretical prediction is free from biases due to choices of distribution, but it neglects prior knowledge of the system e.g., star formation history, age-metallicity relations, and measurements of the metallicity.~One could study the impact of this prior knowledge by imposing them as either relations between the parameters when they are drawn, or as underlying distributions.~In doing so, one should ensure that the errors in such relations/distributions are correctly estimated and fully propagated into the analysis to avoid biasing the results.

As an example of how additional information can be folded into this analysis, we study the additional information imparted when metallicity measurements are incorporated.~One such measurement is reported by \cite{2006ApJ...642..834K}, who measure $Z=0.0091\pm0.0007$.~This measurement was made by fitting non-linear pulsation models to the light curves of LMC Cepheids.~This additional relation can be incorporated into our analysis by drawing $Z$ from a Gaussian distribution with mean $\bar{Z}=0.0091$ and standard deviation $\sigma_Z=0.0091$.~Imposing this prior, we predict the median and interquartile range of $M_I$ to be $M_I = -3.87^{+0.11}_{-0.08}$ in the LMC.~The resulting distribution is shown in the right panel of Figure \ref{fig:iBandDistSM}.~This measurement is also an example of how one must be confident in the relations/distributions chosen since the MC analysis does not provide a method of evaluating their robustness.~In this specific example, there are two major caveats.~First, one is assuming that LMC Cepheids have the same metallicity as LMC stars at the TRGB, which is unlikely.~Second, it is possible that the error bars are underestimated because fitting Cepheid pulsation models to light-curves does not simultaneously vary all parameters and thus does not account for degeneracies.~Indeed \cite{Desmond:2020nde} analyzed LMC Cepheids using an MCMC method similar to the one we present below that simultaneously varies both stellar input physics and nuisance parameters e.g., the strength of turbulent convection.~This method resulted in errors on $Z$ that are an order-of-magnitude larger than those of \cite{2006ApJ...642..834K}.~Thus, while imposing this measurement as a prior has the result of bringing theory closer to observation, it is possible that the prior is too restrictive.~It would be interesting to study the effects of other relations e.g., age-metallicity \cite{2014A&A...563A...5S}.~Our reproduction package provides a convenient starting point for such studies.

\subsection{Comparison of Theoretical and Empirically-Calibrated I Band Magnitudes}

\label{sec:MCMC}

In this section we demonstrate how using ML to emulate the theoretical TRGB $M_I$ can enable comparisons of theory and observation that account for degeneracies by using MCMC sampling to constrain the metallicity in the LMC, NGC 4258, and $\omega$-Centauri, all of which have reported empirical calibrations of $M_I$.~We use the empirical calibrations reported by \cite{2017ApJ...835...28J} for NCG 4258, \cite{2001ApJ...556..635B} for $\omega$-Centauri, and \cite{2017ApJ...835...28J} for the LMC but with a zero point adjustment to the value reported by \cite{2019ApJ...886...61Y}.

The MCMC analysis was performed using the \textit{emcee} package \cite{2013PASP..125..306F} assuming uniform priors on $M$, $Y$, and $Z$ with ranges $0.7{\rm M}_\odot\le M\le2.25 {\rm M}_\odot$, $0.2\le Y\le 0.3$, and $10^{-5}\le Z\le10^{-2}$.~We chose to use uniform priors, which is tantamount to assuming that we have no knowledge of the underlying distribution, for two reasons.~First, we want to test the efficacy of our methodology in the most challenging scenario by exploring whether the MCMC could converge to a localized region of parameter space in the absence of any restrictive priors --- ultimately it did.~Second, we wanted to interpret our results without the influence of the potential caveats associated with imposing priors due to independent measurements or incorporating additional relations discussed above in section \ref{sec:MC}.~The ML classification algorithm was used within the MCMC to assign zero-likelihood to models whose input parameters would give rise to models that burn helium stably in the core or models that are older than the age of the universe.~The log-likelihood function was taken to be Gaussian i.e.,
\begin{multline}
    \ln{\mathcal{L}} = -\frac{1}{2}\left[\left(\frac{M_{I, \textrm{obs}} - \textrm{ML}(\theta)_{I}}{\sigma_{\textrm{tot}}}\right)^2 \right.~ \\ +\left.~\ln(2\pi \sigma_{\textrm{tot}}^2)\vphantom{\sum^N_{i=1}}\right]
    \label{eq:MLE},
\end{multline}
where $M_{I, \textrm{obs}}$ is the observed I Band calibration, ML$(\theta)_{I}$ is the  prediction for $M_I$ from the ML emulator for a given set of parameters $\theta$.~$\sigma_{\textrm{tot}}$ is given by
\begin{equation}
\sigma_{\textrm{tot}}^2 = \sigma_{\textrm{obs}}^2 + \sigma_{\eta}^2,
\end{equation}
where $\sigma_{\textrm{obs}}$ is the error on the empirical I band measurement and $\sigma_\eta$ is the error due to wind loss calculated in section \ref{sec:wind_uncertainties}

\begin{figure}
    \centering
    \includegraphics[scale=0.42]{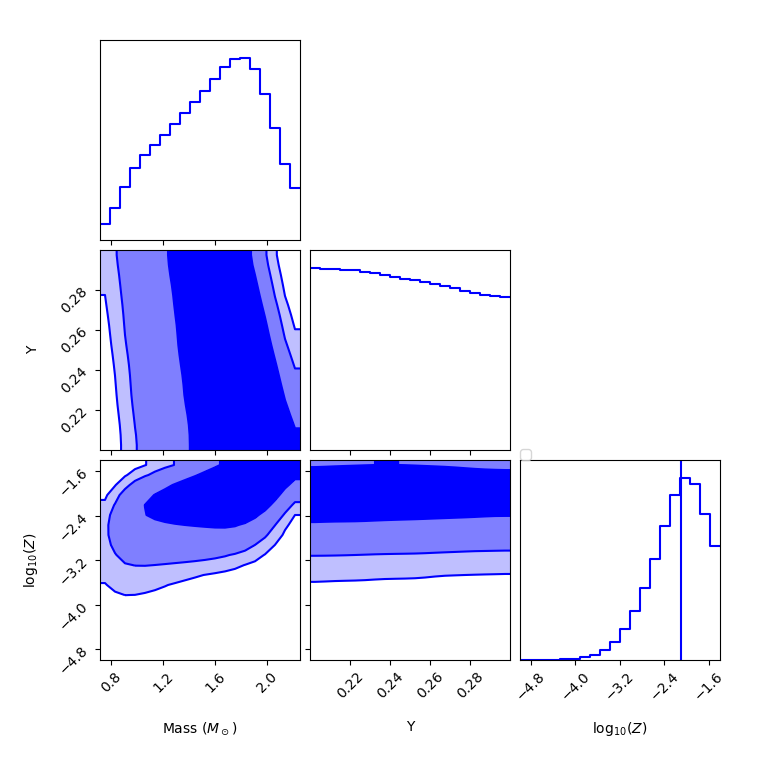}
    \caption{Posterior distributions of each parameter for the $M_I$ calibration in the LMC (Y19).~Both 2D and marginalized posteriors are shown.~The values for the medians and interquartile range are given in \ref{tbl:results}.~The contours represent the 68\%, 95\%, and 99\% confidence intervals.~We show the results for the metallicity in log space as its range spans several orders of magnitude.~The metallicity measurement is consistent with measurements which give $\log_{10}(Z\sim -2.097$}
    \label{fig:corner1}
\end{figure}

\begin{figure}[h!]
    \centering
    \includegraphics[scale=0.42]{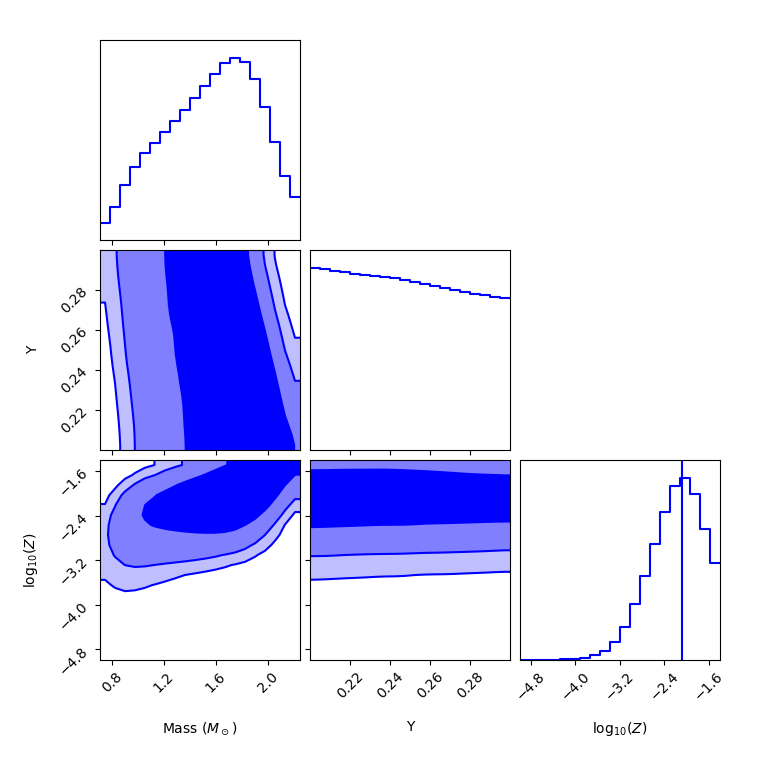}
    \caption{Same as Figure \ref{fig:corner1} but for the $M_I$ calibration in NGC 4258.~This value is also consistent with measurements of NGC 4258 $\log_{10}(Z\sim -2.642$ using the [M/H] metallicity estimate from \cite{new_ngc4258_Z} and \cite{new_ngc4258_Z_2}}
    \label{fig:corner4}
\end{figure}

The error on the machine learning in $M_I$, $\Delta M_I$, was included by randomly drawing values from the distributions given in Figure \ref{fig:mError} and adding them to the values predicted by the ML emulator for each set of parameters $\theta$ sampled.~This is similar to the procedure used by \cite{2019MNRAS.482.1352M}.~These errors are highly subdominant to the errors on the calibration and the bolometric correction.~We determined convergence by demanding that the autocorrelation time $\tau$ was less than 0.1\% the length of the chain and that it had changed by less than 1\% over the previous 10,000 steps \cite{2010CAMCS...5...65G,Sokal}.~In all cases we found that the MCMC converged within 120,000 steps, but we allowed it to continue to 500,000 to ensure that the walker was no longer influenced by its starting location.~We discarded half of the sampler as burn-in.~A corner plot showing the results of the MCMC for Y19 is given in Figure \ref{fig:corner1}.~The corner plots for the other systems we studied are visually similar and are included in the appendix in Figures \ref{fig:corner2}, \ref{fig:corner3}, and \ref{fig:corner4}.

As seen in Table~\ref{tbl:results}, our analysis yields a relatively broad posterior for the mass and helium abundance of the TRGB star(s) in $\omega$-Centauri, while providing a more informative constraint on the metallicity: $\log_{10}(Z) = -2.14^{+0.33}_{-0.27}$, or equivalently $Z = 0.00798^{+0.00733}_{-0.00431}$.~This behavior is expected from the structure of the likelihood function: as shown in Figure~\ref{fig:degeneracy}, a wide range of $M$ and $Y$ values can reproduce the empirical TRGB magnitude at a given $Z$, whereas only a narrow range of metallicities are compatible with the observed calibration.~The resulting constraint on $Z$ arises as a volume effect, reflecting the concentration of viable models in a narrow metallicity range, even as broader ranges of $M$ and $Y$ remain allowed.

Our constraints on the metallicity for all three of our systems are in good agreement with other measurements, within 1$\sigma$ for the LMC and NGC 4258 and 2$\sigma$ or 1$\sigma$ for $\omega$-Centauri depending on the stellar population measured.~~Measurements in the LMC give $\log_{10}(Z\sim -2.097$ (e.g.,\cite{1993A&AS...98..523S, 2006ApJ...642..834K, 10.1007/978-3-540-34136-9_50, 2010A&A...517A..50G, 2013ApJ...768L...6M, 2021MNRAS.507.4752C}); measurements of $\omega$-Centauri give $\log_{10}(Z) \sim -3.201$ calculated with the [Fe/H] metallicity estimate $[{\rm Fe}/{\rm H}] \sim -1.35$ found from spectroscopic metallicity estimates taken from \cite{2010MNRAS.404.1203F, 2009ApJ...694.1498M} and using an $[\alpha /{\rm Fe}] = 0.35$ adopted by \cite{2004A&A...424..199B}.~We calculated the values of metallicity using the method given in \cite{1993ApJ...414..580S} provided by the online calculator from \href{http://www2.astro.puc.cl/pgpuc/FeHcalculator2.php}{PGPUC}.~By extension, our value for $\omega$-Centauri is also consistent within 1$\sigma$ to the spread reported by \cite{2014ApJ...791..107V} (see their table 1, specifically populations 4, 5, and 6).~To convert these [Fe/H] values to Z, we used the same procedure as before.~For NGC 4258 our measurement (converted to $[Z] \approx -0.27$ using a solar metallicity of 0.0132 \cite{solar_z}) is within the errors reported by \cite{new_ngc4258_Z} for their measurement $[Z] = -0.2 \pm 0.1$ and the $-0.29 \pm 0.14$ reported by \cite{new_ngc4258_Z_2} ~Note that the metallicity measurement for $\omega$-Centauri does not include uncertainties, as the values used to compute them were reported without error estimates.~As a result, the quoted level of agreement in terms of $\sigma$ represents a lower bound on the true level of consistency.~A key advantage of our method is that it enables the simultaneous derivation of both measurements and their associated uncertainties.

Furthermore, within errors, all three constraints are compatible with the expectation that the TRGB is determined by metal-poor stars \cite{2000ApJS..128..431F, 2020ApJ...897..106K}.~The error bars are primarily driven by the large errors in the bolometric corrections that are necessary to compute $M_I$ from the output of stellar structure codes.~These are of the order $\Delta M_I\sim 0.1$.~The uncertainty on our constraints could be reduced by imposing informative priors on $Z$ from independent determinations, or by imposing relations between the parameters deriving from e.g., age-metallicity relations \cite{2014A&A...563A...5S}.~Such studies are beyond the scope of the present work, but would make an interesting topic for follow-up investigations.

Turning to the mass, the posteriors are relatively broad.~For $\omega$-Centauri, the peak lies near $1.5M_\odot$, but our model ensemble includes configurations with lower masses and older ages consistent with the system’s known age of $\sim$10--12Gyr \cite{Hughes1999,Stanford2006,2010MNRAS.404.1203F,2024ApJ...977...14C}, with such models appearing within $\sim$2$\sigma$ of the posterior.~This apparent tension reflects a degeneracy between mass, age, and metallicity:~a range of parameter combinations can reproduce the observed TRGB magnitude, especially when no age prior is imposed.

In this work, we have intentionally adopted a prior-independent approach to isolate what the TRGB magnitude alone can constrain.~While metallicity is constrained by this method, mass is not, due to its coupling with other parameters.~Looking ahead, our results suggest that incorporating age and other population-level priors could reduce degeneracies, yielding more precise and physically meaningful constraints.~Our framework is well-suited to such an extension, and we plan to pursue this in future work, particularly as we expand the emulator to include additional stellar parameters.

Finally, the posterior on $Y$ spans the entire prior range, indicating that no meaningful constraint can be placed on the helium abundance from the TRGB magnitude alone.~As shown in Fig.~\ref{fig:degeneracy}, this is due to the fact that models spanning the full prior range in $Y$ remain consistent with the observed TRGB calibration.~This result suggests that future applications of our framework may reduce computational cost and complexity by fixing $Y$ to a fiducial value (thus eliminating it as a free parameter), or by restricting it to a narrower, physically motivated range — for example, bounded below by Big Bang nucleosynthesis constraints and above by chemical evolution models.

\begin{table*}
\centering
\begin{tabular}{|c|c|c|c|c|c|}
    \hline
    Target & Ref.~& $M_I$ & Mass (${\rm M}_\odot$) & $\log_{10}Z$ \\
    \hline
    LMC (Y19) & \cite{2019ApJ...886...61Y} & $-3.958 \pm 0.046$ & $1.612^{+0.231}_{-0.303}$ & $\log_{10}(Z)=-2.098^{+0.388}_{-0.528}$\\
    \hline
    LMC (F20) & \cite{2020ApJ...891...57F} & $-4.047 \pm 0.045$ & $1.536^{+0.261}_{-0.251}$ & $\log_{10}(Z)=-2.167^{+0.404}_{-0.492}$\\
    \hline
    NGC 4258 & \cite{2017ApJ...835...28J} & $-4.016 \pm 0.058$ & $1.579^{+0.230}_{-0.292}$ & $\log_{10}(Z)=-2.146^{+0.400}_{-0.505}$\\
    \hline
    $\omega$-Centauri & \cite{2001ApJ...556..635B} & $-4.012 \pm 0.12$ & $1.583^{+0.229}_{-0.373}$ & $\log_{10}(Z)=-2.143^{+0.401}_{-0.508}$ \\
    \hline
\end{tabular}
\caption{Empirical $M_I$ calibrations, associated error, and the parameter constraints found by the MCMC for each target galaxy.~The median and interquartile range are reported for $M$ and $Z$.~The $Y$ parameter is not included as the posterior is essentially flat and uninformative.~We choose to report the median and interquartile ranges rather than the mean and standard deviation due to the highly non-Gaussian nature of the marginalized posteriors.~Note that the LMC calibration is identical to NGC 4258 but with a shifted zero point.}
\label{tbl:results}
\end{table*}

\section{Discussion}
\label{sec:discussion}

In this section we discuss the limitations of our methodology presented above, and suggest improvements to our preliminary exploration that could ameliorate these in future studies.~We also discuss potential applications of our emulator to other areas of cosmology and astronomy.

\subsection{Limitations of Our Methodology}

The first limitation of our study is that we have fixed the input physics e.g., the opacity, mixing length, nuclear reaction rates, plasma neutrino losses, etc.~Our analysis therefore includes stochastic variations due to environment, winds, and mass but not uncertainties in the input physics.~Fixing the input physics made our procedure computationally tractable and we were able to run 124,844 models in one million CPU hours (using eight cores per model).~As an example, varying five parameters instead of three would require a one million model grid to train the machine learning emulator with acceptable errors.~This would require approximately 16 million CPU hours assuming 16 steps in each parameter.~A second reason we fixed the input physics was to test the efficiency of the ML emulator, which generally degrades with more free parameters.~

We tested the training and validation accuracy of our emulator, and our results indicate that all input physics parameters could be varied without significantly degrading the ML accuracy.~We also tested the accuracy as a function of training grid size and found that as many as 50\% of the models could have been removed before the errors became unacceptable.~This promising result suggests that more parameters could be varied in future studies because fewer grid points per parameter than we initially anticipated are necessary.~We remark that more efficient sampling e.g., Latin hypercube sampling or active learning \cite{2022arXiv220316683A} may further reduce the number of models required.

The second limitation is that we assumed the TRGB $M_I$ is due solely to a single star.~In practice, there are many stars near the TRGB and an edge-detection method (e.g.~\cite{1993ApJ...417..553L}) is used to determine the calibration.~The single star approximation is justified because the peak $M_I$ found using the edge-detection method can reasonably be assumed to be due to the brightest star in the system.~Future work could allow for a more direct comparison by using our ML emulators to stochastically generate mock color-magnitude diagrams and apply the same edge-fitting methods to generate theoretical predictions.~One could then estimate the theoretical error by repeating this process several times.This procedure entails making further assumptions about the underlying distribution of the parameters --- some of which are unknown --- and their associated error, so requires a more comprehensive study that is beyond the scope of the present work.

The final limitation is that our analysis includes statistical errors but not systematic errors.~We have used a single stellar structure code (MESA) and assumed that its predictions are accurate.~Any potential missing physics is not accounted for.~For example, MESA is a one-dimensional code and is missing the physics of three-dimensional processes.~Furthermore, we have made discrete choices for input physics such as boundary conditions, elemental abundances, and numerical solver.~Different choices may result in systematically different $M_I$ predictions.~Finally, we used the WL bolometric corrections but there are alternatives such as MARCS \cite{2008A&A...486..951G} or PHOENIX \cite{2008ApJS..178...89D} that may give systematically different results.~It would be interesting to repeat this work using different stellar structure codes, different choice for input physics, and/or different bolometric corrections and compare the results.

\subsection{Applications to Cosmology and Astronomy}

\subsubsection{Validating Competing Empirical \texorpdfstring{$M_I$}{MI} Calibrations}
The empirical calibration of $M_I$ in the LMC is currently the subject of intense discussion in the context of the Hubble tension \cite{Verde:2019ivm,Abdalla:2022yfr}.~We compared the LMC calibration given by Y19 which gives a value of Hubble's constant, $H_0$, that is consistent with the Cepheid-calibrated distance ladder but in $\sim$5$\sigma$ tension with the value inferred from the CMB \cite{2021arXiv211204510R} with the alternative calibration was reported by F20 that is consistent with the CMB.~Our resulting metallicities ($\log_{10}(Z)=-2.167^{+0.404}_{-0.492}$ and $\log_{10}(Z)=-2.098^{+0.388}_{-0.528}$ for F20 and Y19 respectively) have significant overlap so this method cannot yet be used to validate one value of $H_0$ over the other by demanding consistency with independent measurements of metallicity (e.g., \cite{1993A&AS...98..523S,2006ApJ...642..834K,10.1007/978-3-540-34136-9_50,2010A&A...517A..50G,2013ApJ...768L...6M,2021MNRAS.507.4752C}).~It is possible that the large errors due to the bolometric corrections discussed above resulted in overlapping $Z$ measurements for both LMC calibrations, in which case this validation method would first require the errors on the empirical bolometric corrections to be reduced.

\subsubsection{Constraining Stellar Modeling Parameters}

Our preliminary analysis shows that additional stellar modeling parameters can be varied without significantly degrading emulator accuracy, making it feasible to constrain them using our MCMC framework.~Varying stellar modeling parameters that control the input physics such as the mixing length, radiative opacity, nuclear reaction rates, and plasma neutrino losses --- which are all highly uncertain --- and emulating the results would enable measurements of these parameters to be made using the MCMC methods we have developed here.

In order to advance this program and achieve its goals, it is essential to mitigate the degeneracies shown in Fig.~\ref{fig:degeneracy}, as well as others that will inevitably arise when additional parameters are varied.~Reducing these degeneracies would narrow the posteriors and enable more meaningful constraints.~We have already discussed several strategies for doing so, such as imposing age priors (when available), which can restrict the allowed mass range, or fixing $Y$ --- or limiting it to a narrower, physically motivated prior range.

The constraining power can also be improved by increasing the amount of observational data.~For example, TRGB measurements in multiple bands --- such as the $I$, $J$, $H$, and $K$ bands --- can provide additional constraints and help break parameter degeneracies by leveraging the distinct spectral sensitivities of each band to different stellar properties;~such calibrations exist for many systems \cite{2018AJ....156..278G}.~Another strategy is to jointly fit multiple systems simultaneously.~This increases statistical power and enables one to disentangle stochastic parameters --- such as mass, metallicity, helium abundance, and wind loss, which vary from system to system --- from universal parameters like nuclear reaction rates, opacity, and neutrino energy loss rates, which should be identical across all systems.~This approach enhances inference by separating system-specific effects from universal physics.~For instance, \cite{2018AJ....156..278G} report $M_I$ for 19 systems in the Small and Large Magellanic Clouds, including magnitudes in the $J$, $H$, and $K$ bands.~Similarly, \cite{2010MNRAS.404.1203F} provide measurements for 93 Milky Way globular clusters, which could offer a wealth of additional data, provided $M_{I,\mathrm{TRGB}}$ measurements are available.

\section{Summary and Conclusion}
\label{sec:conc}

In this work we have presented a methodology to explore the effects of environment and input physics on the TRGB $M_I$.~We simulated a grid of 124,844 stellar models with varying mass, initial helium abundance, and metallicity and used this to train a machine learning emulator to predict $M_I$ as a function of these parameters.~We used this to perform a Monte Carlo analysis where we randomly drew 5,000,000 sets of parameters (assuming them to be independently uniformly distributed) to derive a theoretical distribution for $M_I$.~We predict $M_I = -3.87^{+0.11}_{-0.08}$ in the LMC.~It is likely the true uncertainty is larger since we fixed the input physics.~This paper is accompanied by a reproduction package that contains our emulator, and the  grid of models used to train it \cite{dennis_mitchell_t_2022_7895972}.

Our emulator enables direct comparison between theoretical and empirical I-band TRGB calibrations through Markov Chain Monte Carlo analysis.~Since each MCMC step requires evaluation in seconds or less to remain tractable, performing full stellar structure simulations (which can take hours per model) is infeasible.~In contrast, our emulator evaluates in milliseconds, making such analyses computationally efficient and enabling exploration of high-dimensional parameter spaces.~Our results are summarized in Tabl~\ref{tbl:results}.

We applied empirical calibrations from the LMC (F20 and Y19), NGC~4258, and $\omega$-Centauri to constrain the metallicities of these systems, finding good agreement with published values (within 1–2$\sigma$).~We also obtained constraints on stellar mass, though the posteriors are broader.~In the case of $\omega$-Centauri, for example, the mass posterior peaks near $1.5\,M_\odot$, which is inconsistent with its known age of $\sim$10–12 Gyr, but lower mass, older models that align with this age appear within 2$\sigma$ of the distribution.~As discussed in Section~\ref{sec:MCMC}, incorporating age priors informed by stellar population studies could help break these degeneracies and yield tighter constraints.~We plan to implement this in future extensions of our framework.

We note that our results should be interpreted as constraints on the properties of the brightest TRGB star(s) in each system, rather than on the underlying stellar population as a whole.~Our methodology is motivated by the fact that the TRGB magnitude is observationally defined via edge-detection methods applied to the luminosity function in color-magnitude diagrams, which identify a sharp cutoff corresponding to the brightest red giant stars undergoing the helium flash.~While stellar systems contain a range of masses, metallicities, and ages, the TRGB calibration is effectively set by the properties of a narrow subset of stars near this luminosity limit \cite{1993ApJ...417..553L, Ivans:2001qp, Sandquist:2004wf, 2013A&A...558A..12V}.~Our analysis focuses on this subset, yielding robust constraints on the stellar parameters most relevant for determining the observed TRGB magnitude.~A future extension of this work could involve simulating full stellar populations and applying the same edge-detection method directly to synthetic color-magnitude diagrams, enabling marginalization over population-level distributions.~We have outlined how such an approach could be implemented in Sections~\ref{sec:newBC} and~\ref{sec:MC}.~Crucially, our machine learning emulator is designed to make such an extension computationally feasible, as it allows rapid prediction of TRGB properties across a high-dimensional parameter space.~Once additional stellar parameters are included and emulated, our framework will be ideally suited for full-population modeling of the TRGB, and we plan to explore this in future work.

Our results are subject to several limitations discussed in section \ref{sec:discussion}.~If these can be overcome then our methodology could be used to measure stellar modeling parameters, or for validating competing empirical $M_I$ calibrations to the LMC that are the source of much discussion in the context of the Hubble tension.

We conclude by noting that the techniques we have introduced here are applicable to other areas of stellar astronomy.~Possible applications include emulating stellar pulsation codes to constrain Cepheid and RR Lyrae properties using MCMC (particularly for such objects in detached eclipsing binaries \cite{Desmond:2020nde}), and emulating main-sequence models for use in isochrone fitting.

\section{Acknowledgements}

We thank Adrian Ayala, Aaron Dotter, Robert Farmer, Frank Timmes, and the wider MESA community for answering our MESA-related questions.~We are grateful for conversations with Eric J.~Baxter, Djuna Croon,  Noah Franz, Marco Gatti, Dan Hey, Esther Hu, Jason Kumar, Danny Marfatia, Marco Raveri, Xerxes Tata, Brent Tully, Guy Worthey, Dan Huber, and Jen van Saders.~We are especially thankful to David Rubin for providing detailed comments and answering our many questions, and to David Schanzenbach for his assistance with using the University of Hawai\okina i  supercomputer MANA.~

Our simulations were run on the University of Hawai‘i’s high-performance supercomputer MANA.~The technical support and advanced computing resources from University of Hawai‘i Information Technology Services – Cyberinfrastructure, funded in part by the National Science Foundation MRI award \#1920304, are gratefully
acknowledged.

\section*{Data Availability}
All data used in this work are publicly available in our reproduction package \cite{dennis_mitchell_t_2022_7895972}.

\section*{Software}
MESA version 12778, MESASDK version 20200325, Worthey and Lee Bolometric Correction Code \cite{2011ApJS..193....1W}, \texttt{Python3} version 3.8.10 \cite{python3}, \texttt{NumPy} version 1.22.3 \cite{2020Natur.585..357H}, \texttt{Pandas} version 1.4.3 \cite{pandasDataStructure, pandasSoftware}, \texttt{Matplotlib} version 3.5.1 \cite{2007CSE.....9...90H}, \texttt{Tensorflow} version 2.4.1 \cite{tensorflow2015-whitepaper, chollet2015keras}, \texttt{scikit-learn} \cite{scikit-learn}, \texttt{corner} version 2.2.1 \cite{corner}, \texttt{emcee} version 3.1.2 \cite{2013PASP..125..306F}, \texttt{imbalanced-learn} version 0.9.0 \citep{imblearn}.

\section*{appendix}
In this section we show the corner plots for our MCMC analyses of the TRGB $M_I$ empirical calibrations in the LMC (F20) $\omega$-Centauri, and NGC 4258.~See section \ref{sec:MCMC} for more details.

\begin{figure}[h!]
    \centering
    \includegraphics[scale=0.42]{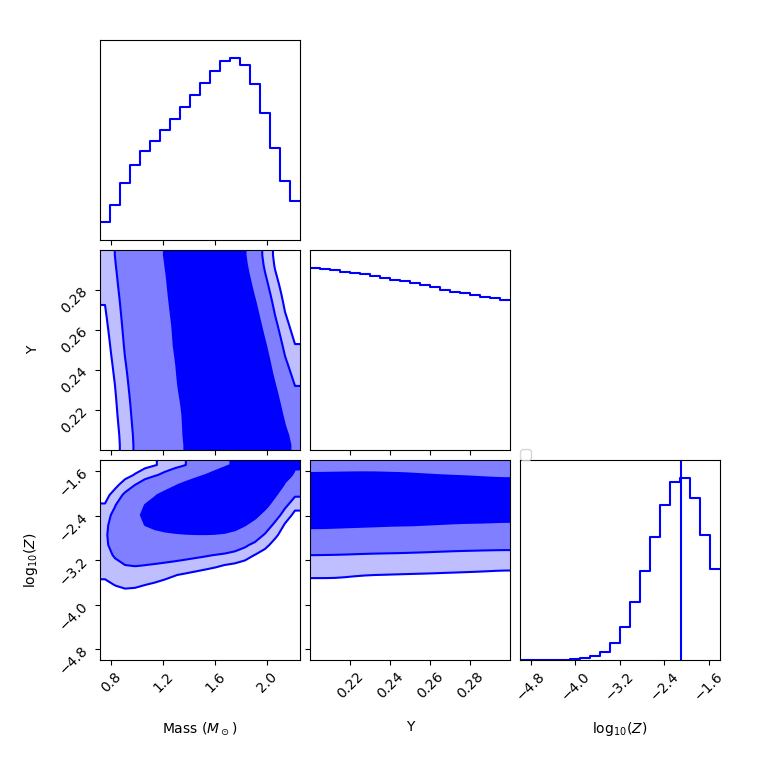}
    \caption{Same as Figure \ref{fig:corner1} but for the other $M_I$ calibration in the LMC (F20).~This value is consistent with measurements which give $\log_{10}(Z \sim -2.097$ (e.g., \cite{1993A&AS...98..523S,2006ApJ...642..834K,10.1007/978-3-540-34136-9_50,2010A&A...517A..50G, 2013ApJ...768L...6M,2021MNRAS.507.4752C}).}
    \label{fig:corner2}
\end{figure}

\begin{figure}[h!]
    \centering
    \includegraphics[scale=0.42]{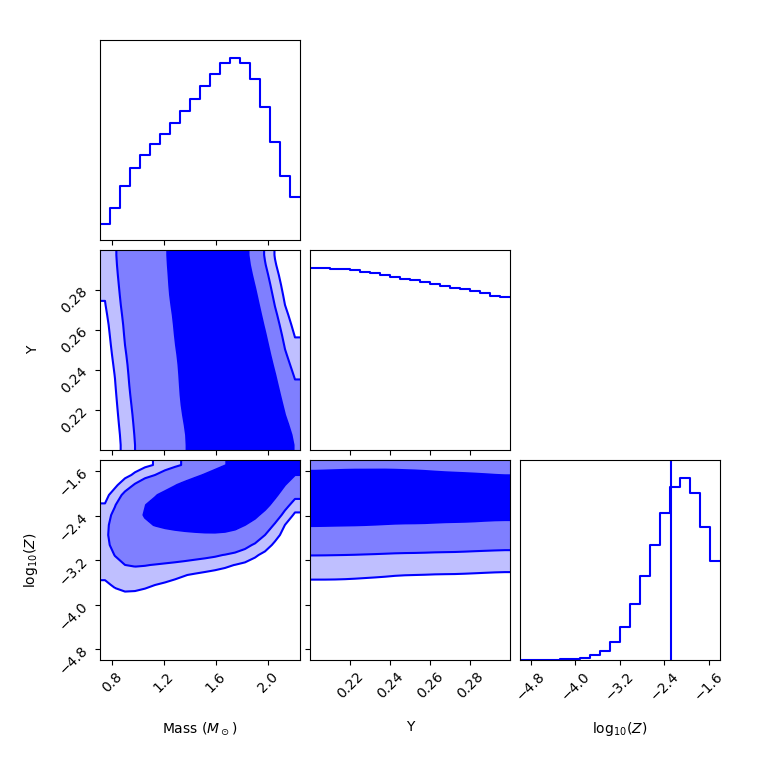}
    \caption{Same as Figure \ref{fig:corner1} but for the $M_I$ calibration in $\omega$-Centauri.~This value is also consistent with measurement of $\omega$-Centauri which gives $\log_{10}(Z\sim -2.28$ calculated with the from [Fe/H] metallicity measurement from \cite{2014ApJ...791..107V}'s population 5.}
    \label{fig:corner3}
\end{figure}

\clearpage
\bibliography{main}

\begin{thebibliography}{}
\makeatletter
\relax
\def\mn@urlcharsother{\let\do\@makeother \do\$\do\&\do\#\do\^\do\_\do\%\do\~}
\def\mn@doi{\begingroup\mn@urlcharsother \@ifnextchar [ {\mn@doi@}
  {\mn@doi@[]}}
\def\mn@doi@[#1]#2{\def\@tempa{#1}\ifx\@tempa\@empty \href
  {http://dx.doi.org/#2} {doi:#2}\else \href {http://dx.doi.org/#2} {#1}\fi
  \endgroup}
\def\mn@eprint#1#2{\mn@eprint@#1:#2::\@nil}
\def\mn@eprint@arXiv#1{\href {http://arxiv.org/abs/#1} {{\tt arXiv:#1}}}
\def\mn@eprint@dblp#1{\href {http://dblp.uni-trier.de/rec/bibtex/#1.xml}
  {dblp:#1}}
\def\mn@eprint@#1:#2:#3:#4\@nil{\def\@tempa {#1}\def\@tempb {#2}\def\@tempc
  {#3}\ifx \@tempc \@empty \let \@tempc \@tempb \let \@tempb \@tempa \fi \ifx
  \@tempb \@empty \def\@tempb {arXiv}\fi \@ifundefined
  {mn@eprint@\@tempb}{\@tempb:\@tempc}{\expandafter \expandafter \csname
  mn@eprint@\@tempb\endcsname \expandafter{\@tempc}}}

\bibitem[\protect\citeauthoryear{Abadi et~al.,}{Abadi
  et~al.}{2015}]{tensorflow2015-whitepaper}
Abadi M.,  et~al., 2015, {TensorFlow}: Large-Scale Machine Learning on
  Heterogeneous Systems, \url {https://www.tensorflow.org/}

\bibitem[\protect\citeauthoryear{Abdalla et~al.}{Abdalla
  et~al.}{2022}]{Abdalla:2022yfr}
Abdalla E.,  et~al., 2022, \mn@doi [JHEAp] {10.1016/j.jheap.2022.04.002}, 34,
  49

\bibitem[\protect\citeauthoryear{{Anand}, {Tully}, {Rizzi}, {Riess}  \&
  {Yuan}}{{Anand} et~al.}{2022}]{2022ApJ...932...15A}
{Anand} G.~S.,  {Tully} R.~B.,  {Rizzi} L.,  {Riess} A.~G.,   {Yuan} W.,  2022,
  \mn@doi [\apj] {10.3847/1538-4357/ac68df}, \href
  {https://ui.adsabs.harvard.edu/abs/2022ApJ...932...15A} {932, 15}

\bibitem[\protect\citeauthoryear{{Anderson}, {Koblischke}  \&
  {Eyer}}{{Anderson} et~al.}{2023}]{2023arXiv230304790A}
{Anderson} R.~I.,  {Koblischke} N.~W.,   {Eyer} L.,  2023, \mn@doi [arXiv
  e-prints] {10.48550/arXiv.2303.04790}, \href
  {https://ui.adsabs.harvard.edu/abs/2023arXiv230304790A} {p. arXiv:2303.04790}

\bibitem[\protect\citeauthoryear{Angulo et~al.}{Angulo
  et~al.}{1999}]{Angulo:1999zz}
Angulo C.,  et~al., 1999, \mn@doi [Nucl. Phys. A]
  {10.1016/S0375-9474(99)00030-5}, 656, 3

\bibitem[\protect\citeauthoryear{{Bellazzini}, {Ferraro}  \&
  {Pancino}}{{Bellazzini} et~al.}{2001}]{2001ApJ...556..635B}
{Bellazzini} M.,  {Ferraro} F.~R.,   {Pancino} E.,  2001, \mn@doi [\apj]
  {10.1086/321613}, \href
  {https://ui.adsabs.harvard.edu/abs/2001ApJ...556..635B} {556, 635}

\bibitem[\protect\citeauthoryear{{Bellazzini}, {Ferraro}, {Sollima}, {Pancino}
  \& {Origlia}}{{Bellazzini} et~al.}{2004}]{2004A&A...424..199B}
{Bellazzini} M.,  {Ferraro} F.~R.,  {Sollima} A.,  {Pancino} E.,   {Origlia}
  L.,  2004, \mn@doi [\aap] {10.1051/0004-6361:20035910}, \href
  {https://ui.adsabs.harvard.edu/abs/2004A&A...424..199B} {424, 199}

\bibitem[\protect\citeauthoryear{{Bellinger}, {Angelou}, {Hekker}, {Basu},
  {Ball}  \& {Guggenberger}}{{Bellinger} et~al.}{2016}]{2016ApJ...830...31B}
{Bellinger} E.~P.,  {Angelou} G.~C.,  {Hekker} S.,  {Basu} S.,  {Ball} W.~H.,
  {Guggenberger} E.,  2016, \mn@doi [\apj] {10.3847/0004-637X/830/1/31}, \href
  {https://ui.adsabs.harvard.edu/abs/2016ApJ...830...31B} {830, 31}

\bibitem[\protect\citeauthoryear{{Bellinger}, {Angelou}, {Hekker}, {Basu},
  {Ball}  \& {Guggenberger}}{{Bellinger} et~al.}{2017}]{2017EPJWC.16005003B}
{Bellinger} E.~P.,  {Angelou} G.~C.,  {Hekker} S.,  {Basu} S.,  {Ball} W.~H.,
  {Guggenberger} E.,  2017, in European Physical Journal Web of Conferences. p.
  05003 (\mn@eprint {arXiv} {1705.06759}),
  \mn@doi{10.1051/epjconf/201716005003}

\bibitem[\protect\citeauthoryear{{Betancourt}}{{Betancourt}}{2017}]{2017arXiv170102434B}
{Betancourt} M.,  2017, \mn@doi [arXiv e-prints] {10.48550/arXiv.1701.02434},
  \href {https://ui.adsabs.harvard.edu/abs/2017arXiv170102434B} {p.
  arXiv:1701.02434}

\bibitem[\protect\citeauthoryear{{Betancourt}, {Byrne}, {Livingstone}  \&
  {Girolami}}{{Betancourt} et~al.}{2014}]{2014arXiv1410.5110B}
{Betancourt} M.~J.,  {Byrne} S.,  {Livingstone} S.,   {Girolami} M.,  2014,
  \mn@doi [arXiv e-prints] {10.48550/arXiv.1410.5110}, \href
  {https://ui.adsabs.harvard.edu/abs/2014arXiv1410.5110B} {p. arXiv:1410.5110}

\bibitem[\protect\citeauthoryear{{Bresolin}}{{Bresolin}}{2011}]{new_ngc4258_Z_2}
{Bresolin} F.,  2011, \mn@doi [\apj] {10.1088/0004-637X/729/1/56}, \href
  {https://ui.adsabs.harvard.edu/abs/2011ApJ...729...56B} {729, 56}

\bibitem[\protect\citeauthoryear{{Chahrour} \& {Wells}}{{Chahrour} \&
  {Wells}}{2022}]{2022ScPP...12..187C}
{Chahrour} I.,  {Wells} J.,  2022, \mn@doi [SciPost Physics]
  {10.21468/SciPostPhys.12.6.187}, \href
  {https://ui.adsabs.harvard.edu/abs/2022ScPP...12..187C} {12, 187}

\bibitem[\protect\citeauthoryear{{Chawla}, {Bowyer}, {Hall}  \&
  {Kegelmeyer}}{{Chawla} et~al.}{2011}]{2011arXiv1106.1813C}
{Chawla} N.~V.,  {Bowyer} K.~W.,  {Hall} L.~O.,   {Kegelmeyer} W.~P.,  2011,
  Journal of Artificial Intelligence Research, \href
  {https://ui.adsabs.harvard.edu/abs/2011arXiv1106.1813C} {p. arXiv:1106.1813}

\bibitem[\protect\citeauthoryear{Chollet et~al.}{Chollet
  et~al.}{2015}]{chollet2015keras}
Chollet F.,  et~al., 2015, Keras, \url{https://keras.io}

\bibitem[\protect\citeauthoryear{{Choudhury} et~al.,}{{Choudhury}
  et~al.}{2021}]{2021MNRAS.507.4752C}
{Choudhury} S.,  et~al., 2021, \mn@doi [\mnras] {10.1093/mnras/stab2446}, \href
  {https://ui.adsabs.harvard.edu/abs/2021MNRAS.507.4752C} {507, 4752}

\bibitem[\protect\citeauthoryear{{Clontz} et~al.,}{{Clontz}
  et~al.}{2024}]{2024ApJ...977...14C}
{Clontz} C.,  et~al., 2024, \mn@doi [\apj] {10.3847/1538-4357/ad8621}, \href
  {https://ui.adsabs.harvard.edu/abs/2024ApJ...977...14C} {977, 14}

\bibitem[\protect\citeauthoryear{{Cyburt} et~al.,}{{Cyburt}
  et~al.}{2010}]{2010ApJS..189..240C}
{Cyburt} R.~H.,  et~al., 2010, \mn@doi [\apjs] {10.1088/0067-0049/189/1/240},
  \href {https://ui.adsabs.harvard.edu/abs/2010ApJS..189..240C} {189, 240}

\bibitem[\protect\citeauthoryear{Dennis \& Sakstein}{Dennis \&
  Sakstein}{2022}]{dennis_mitchell_t_2022_7895972}
Dennis M.~T.,  Sakstein J.,  2022, Machine Learning the Tip of the Red Giant
  Branch, \mn@doi{10.5281/zenodo.7895972}, \url
  {https://doi.org/10.5281/zenodo.7895972}

\bibitem[\protect\citeauthoryear{Desmond, Sakstein  \& Jain}{Desmond
  et~al.}{2021}]{Desmond:2020nde}
Desmond H.,  Sakstein J.,   Jain B.,  2021, \mn@doi [Phys. Rev. D]
  {10.1103/PhysRevD.103.024028}, 103, 024028

\bibitem[\protect\citeauthoryear{{Dotter}, {Chaboyer}, {Jevremovi{\'c}},
  {Kostov}, {Baron}  \& {Ferguson}}{{Dotter}
  et~al.}{2008}]{2008ApJS..178...89D}
{Dotter} A.,  {Chaboyer} B.,  {Jevremovi{\'c}} D.,  {Kostov} V.,  {Baron} E.,
  {Ferguson} J.~W.,  2008, \mn@doi [\apjs] {10.1086/589654}, \href
  {https://ui.adsabs.harvard.edu/abs/2008ApJS..178...89D} {178, 89}

\bibitem[\protect\citeauthoryear{{Ferrarese} et~al.,}{{Ferrarese}
  et~al.}{2000}]{2000ApJS..128..431F}
{Ferrarese} L.,  et~al., 2000, \mn@doi [\apjs] {10.1086/313391}, \href
  {https://ui.adsabs.harvard.edu/abs/2000ApJS..128..431F} {128, 431}

\bibitem[\protect\citeauthoryear{{Forbes} \& {Bridges}}{{Forbes} \&
  {Bridges}}{2010}]{2010MNRAS.404.1203F}
{Forbes} D.~A.,  {Bridges} T.,  2010, \mn@doi [\mnras]
  {10.1111/j.1365-2966.2010.16373.x}, \href
  {https://ui.adsabs.harvard.edu/abs/2010MNRAS.404.1203F} {404, 1203}

\bibitem[\protect\citeauthoryear{Foreman-Mackey}{Foreman-Mackey}{2016}]{corner}
Foreman-Mackey D.,  2016, \mn@doi [The Journal of Open Source Software]
  {10.21105/joss.00024}, 1, 24

\bibitem[\protect\citeauthoryear{{Foreman-Mackey}, {Hogg}, {Lang}  \&
  {Goodman}}{{Foreman-Mackey} et~al.}{2013}]{2013PASP..125..306F}
{Foreman-Mackey} D.,  {Hogg} D.~W.,  {Lang} D.,   {Goodman} J.,  2013, \mn@doi
  [\pasp] {10.1086/670067}, \href
  {https://ui.adsabs.harvard.edu/abs/2013PASP..125..306F} {125, 306}

\bibitem[\protect\citeauthoryear{{Freedman} et~al.,}{{Freedman}
  et~al.}{2019}]{2019ApJ...882...34F}
{Freedman} W.~L.,  et~al., 2019, \mn@doi [\apj] {10.3847/1538-4357/ab2f73},
  \href {https://ui.adsabs.harvard.edu/abs/2019ApJ...882...34F} {882, 34}

\bibitem[\protect\citeauthoryear{{Freedman} et~al.,}{{Freedman}
  et~al.}{2020}]{2020ApJ...891...57F}
{Freedman} W.~L.,  et~al., 2020, \mn@doi [\apj] {10.3847/1538-4357/ab7339},
  \href {https://ui.adsabs.harvard.edu/abs/2020ApJ...891...57F} {891, 57}

\bibitem[\protect\citeauthoryear{{Glatt}, {Grebel}  \& {Koch}}{{Glatt}
  et~al.}{2010}]{2010A&A...517A..50G}
{Glatt} K.,  {Grebel} E.~K.,   {Koch} A.,  2010, \mn@doi [\aap]
  {10.1051/0004-6361/201014187}, \href
  {https://ui.adsabs.harvard.edu/abs/2010A&A...517A..50G} {517, A50}

\bibitem[\protect\citeauthoryear{{Gonz{\'a}lez-L{\'o}pezlira}
  et~al.,}{{Gonz{\'a}lez-L{\'o}pezlira} et~al.}{2019}]{new_ngc4258_Z}
{Gonz{\'a}lez-L{\'o}pezlira} R.~A.,  et~al., 2019, \mn@doi [\apj]
  {10.3847/1538-4357/ab113a}, \href
  {https://ui.adsabs.harvard.edu/abs/2019ApJ...876...39G} {876, 39}

\bibitem[\protect\citeauthoryear{{Goodman} \& {Weare}}{{Goodman} \&
  {Weare}}{2010}]{2010CAMCS...5...65G}
{Goodman} J.,  {Weare} J.,  2010, \mn@doi [Communications in Applied
  Mathematics and Computational Science] {10.2140/camcos.2010.5.65}, \href
  {https://ui.adsabs.harvard.edu/abs/2010CAMCS...5...65G} {5, 65}

\bibitem[\protect\citeauthoryear{{G{\'o}rski} et~al.,}{{G{\'o}rski}
  et~al.}{2018}]{2018AJ....156..278G}
{G{\'o}rski} M.,  et~al., 2018, \mn@doi [\aj] {10.3847/1538-3881/aaeacb}, \href
  {https://ui.adsabs.harvard.edu/abs/2018AJ....156..278G} {156, 278}

\bibitem[\protect\citeauthoryear{{Grevesse} \& {Sauval}}{{Grevesse} \&
  {Sauval}}{1998}]{1998SSRv...85..161G}
{Grevesse} N.,  {Sauval} A.~J.,  1998, \mn@doi [\ssr]
  {10.1023/A:1005161325181}, \href
  {https://ui.adsabs.harvard.edu/abs/1998SSRv...85..161G} {85, 161}

\bibitem[\protect\citeauthoryear{{Grevesse}, {Asplund}, {Sauval}  \&
  {Scott}}{{Grevesse} et~al.}{2012}]{solar_z}
{Grevesse} N.,  {Asplund} M.,  {Sauval} A.~J.,   {Scott} P.,  2012, in
  {Shibahashi} H.,  {Takata} M.,   {Lynas-Gray} A.~E.,  eds,  Astronomical
  Society of the Pacific Conference Series Vol. 462, Progress in Solar/Stellar
  Physics with Helio- and Asteroseismology. p.~41

\bibitem[\protect\citeauthoryear{{Grinsztajn}, {Oyallon}  \&
  {Varoquaux}}{{Grinsztajn} et~al.}{2022}]{2022arXiv220708815G}
{Grinsztajn} L.,  {Oyallon} E.,   {Varoquaux} G.,  2022, \mn@doi [arXiv
  e-prints] {10.48550/arXiv.2207.08815}, \href
  {https://ui.adsabs.harvard.edu/abs/2022arXiv220708815G} {p. arXiv:2207.08815}

\bibitem[\protect\citeauthoryear{{Gustafsson}, {Edvardsson}, {Eriksson},
  {J{\o}rgensen}, {Nordlund}  \& {Plez}}{{Gustafsson}
  et~al.}{2008}]{2008A&A...486..951G}
{Gustafsson} B.,  {Edvardsson} B.,  {Eriksson} K.,  {J{\o}rgensen} U.~G.,
  {Nordlund} {\r{A}}.,   {Plez} B.,  2008, \mn@doi [\aap]
  {10.1051/0004-6361:200809724}, \href
  {https://ui.adsabs.harvard.edu/abs/2008A&A...486..951G} {486, 951}

\bibitem[\protect\citeauthoryear{{Hansen}, {Kawaler}  \& {Trimble}}{{Hansen}
  et~al.}{2004}]{2004sipp.book.....H}
{Hansen} C.~J.,  {Kawaler} S.~D.,   {Trimble} V.,  2004, {Stellar interiors :
  physical principles, structure, and evolution}

\bibitem[\protect\citeauthoryear{{Harris} et~al.,}{{Harris}
  et~al.}{2020}]{2020Natur.585..357H}
{Harris} C.~R.,  et~al., 2020, \mn@doi [\nat] {10.1038/s41586-020-2649-2},
  \href {https://ui.adsabs.harvard.edu/abs/2020Natur.585..357H} {585, 357}

\bibitem[\protect\citeauthoryear{Hughes \& Wallerstein}{Hughes \&
  Wallerstein}{1999}]{Hughes1999}
Hughes J.,  Wallerstein G.,  1999, \mn@doi [arXiv e-prints]
  {10.48550/arXiv.astro-ph/9912291}, \href
  {https://ui.adsabs.harvard.edu/abs/1999astro.ph.12291H} {}

\bibitem[\protect\citeauthoryear{{Hunter}}{{Hunter}}{2007}]{2007CSE.....9...90H}
{Hunter} J.~D.,  2007, \mn@doi [Computing in Science and Engineering]
  {10.1109/MCSE.2007.55}, \href
  {https://ui.adsabs.harvard.edu/abs/2007CSE.....9...90H} {9, 90}

\bibitem[\protect\citeauthoryear{Ivans, Kraft, Sneden, Smith, Rich  \&
  Shetrone}{Ivans et~al.}{2001}]{Ivans:2001qp}
Ivans I.~I.,  Kraft R.~P.,  Sneden C.,  Smith G.~H.,  Rich R.~M.,   Shetrone
  M.,  2001, \mn@doi [Astron. J.] {10.1086/322108}, 122, 1438

\bibitem[\protect\citeauthoryear{{Jang} \& {Lee}}{{Jang} \&
  {Lee}}{2017}]{2017ApJ...835...28J}
{Jang} I.~S.,  {Lee} M.~G.,  2017, \mn@doi [\apj] {10.3847/1538-4357/835/1/28},
  \href {https://ui.adsabs.harvard.edu/abs/2017ApJ...835...28J} {835, 28}

\bibitem[\protect\citeauthoryear{{Jermyn} et~al.,}{{Jermyn}
  et~al.}{2022}]{2022arXiv220803651J}
{Jermyn} A.~S.,  et~al., 2022, arXiv e-prints, \href
  {https://ui.adsabs.harvard.edu/abs/2022arXiv220803651J} {p. arXiv:2208.03651}

\bibitem[\protect\citeauthoryear{Jimenez, Thejll, Jorgensen, MacDonald  \&
  Pagel}{Jimenez et~al.}{1996}]{Jimenez:1996at}
Jimenez R.,  Thejll P.,  Jorgensen U.,  MacDonald J.,   Pagel B.,  1996,
  \mn@doi [Mon. Not. Roy. Astron. Soc.] {10.1093/mnras/282.3.926}, 282, 926

\bibitem[\protect\citeauthoryear{{Jorgensen} \& {Thejll}}{{Jorgensen} \&
  {Thejll}}{1993}]{1993A&A...272..255J}
{Jorgensen} U.~G.,  {Thejll} P.,  1993, \aap, \href
  {https://ui.adsabs.harvard.edu/abs/1993A&A...272..255J} {272, 255}

\bibitem[\protect\citeauthoryear{{Kang}, {Kim}, {Lee}  \& {Jang}}{{Kang}
  et~al.}{2020}]{2020ApJ...897..106K}
{Kang} J.,  {Kim} Y.~J.,  {Lee} M.~G.,   {Jang} I.~S.,  2020, \mn@doi [\apj]
  {10.3847/1538-4357/ab94ba}, \href
  {https://ui.adsabs.harvard.edu/abs/2020ApJ...897..106K} {897, 106}

\bibitem[\protect\citeauthoryear{{Keller} \& {Wood}}{{Keller} \&
  {Wood}}{2006}]{2006ApJ...642..834K}
{Keller} S.~C.,  {Wood} P.~R.,  2006, \mn@doi [\apj] {10.1086/501115}, \href
  {https://ui.adsabs.harvard.edu/abs/2006ApJ...642..834K} {642, 834}

\bibitem[\protect\citeauthoryear{{Kingma} \& {Ba}}{{Kingma} \&
  {Ba}}{2014}]{2014arXiv1412.6980K}
{Kingma} D.~P.,  {Ba} J.,  2014, arXiv e-prints, \href
  {https://ui.adsabs.harvard.edu/abs/2014arXiv1412.6980K} {p. arXiv:1412.6980}

\bibitem[\protect\citeauthoryear{{Kippenhahn}, {Weigert}  \&
  {Weiss}}{{Kippenhahn} et~al.}{2013}]{2013sse..book.....K}
{Kippenhahn} R.,  {Weigert} A.,   {Weiss} A.,  2013, {Stellar Structure and
  Evolution}, \mn@doi{10.1007/978-3-642-30304-3.
}

\bibitem[\protect\citeauthoryear{Ko{\l}aczkowski \& Pigulski}{Ko{\l}aczkowski
  \& Pigulski}{2006}]{10.1007/978-3-540-34136-9_50}
Ko{\l}aczkowski Z.,  Pigulski A.,  2006, in Randich S.,  Pasquini L.,  eds,
  Chemical Abundances and Mixing in Stars in the Milky Way and its Satellites.
  Springer Berlin Heidelberg, Berlin, Heidelberg, pp 136--137

\bibitem[\protect\citeauthoryear{{Lee}, {Freedman}  \& {Madore}}{{Lee}
  et~al.}{1993}]{1993ApJ...417..553L}
{Lee} M.~G.,  {Freedman} W.~L.,   {Madore} B.~F.,  1993, \mn@doi [\apj]
  {10.1086/173334}, \href
  {https://ui.adsabs.harvard.edu/abs/1993ApJ...417..553L} {417, 553}

\bibitem[\protect\citeauthoryear{Lema{{\^{i}}}tre, Nogueira  \&
  Aridas}{Lema{{\^{i}}}tre et~al.}{2017}]{imblearn}
Lema{{\^{i}}}tre G.,  Nogueira F.,   Aridas C.~K.,  2017, Journal of Machine
  Learning Research, 18, 1

\bibitem[\protect\citeauthoryear{{Liashchynskyi} \&
  {Liashchynskyi}}{{Liashchynskyi} \&
  {Liashchynskyi}}{2019}]{2019arXiv191206059L}
{Liashchynskyi} P.,  {Liashchynskyi} P.,  2019, arXiv e-prints, \href
  {https://ui.adsabs.harvard.edu/abs/2019arXiv191206059L} {p. arXiv:1912.06059}

\bibitem[\protect\citeauthoryear{{Marconi} et~al.,}{{Marconi}
  et~al.}{2013}]{2013ApJ...768L...6M}
{Marconi} M.,  et~al., 2013, \mn@doi [\apjl] {10.1088/2041-8205/768/1/L6},
  \href {https://ui.adsabs.harvard.edu/abs/2013ApJ...768L...6M} {768, L6}

\bibitem[\protect\citeauthoryear{{Mar{\'\i}n-Franch}
  et~al.,}{{Mar{\'\i}n-Franch} et~al.}{2009}]{2009ApJ...694.1498M}
{Mar{\'\i}n-Franch} A.,  et~al., 2009, \mn@doi [\apj]
  {10.1088/0004-637X/694/2/1498}, \href
  {https://ui.adsabs.harvard.edu/abs/2009ApJ...694.1498M} {694, 1498}

\bibitem[\protect\citeauthoryear{{McClintock} et~al.,}{{McClintock}
  et~al.}{2019}]{2019MNRAS.482.1352M}
{McClintock} T.,  et~al., 2019, \mn@doi [\mnras] {10.1093/mnras/sty2711}, \href
  {https://ui.adsabs.harvard.edu/abs/2019MNRAS.482.1352M} {482, 1352}

\bibitem[\protect\citeauthoryear{{Mukherjee}, {Khare}  \& {Verma}}{{Mukherjee}
  et~al.}{2019}]{2019arXiv191011605M}
{Mukherjee} K.,  {Khare} A.,   {Verma} A.,  2019, \mn@doi [arXiv e-prints]
  {10.48550/arXiv.1910.11605}, \href
  {https://ui.adsabs.harvard.edu/abs/2019arXiv191011605M} {p. arXiv:1910.11605}

\bibitem[\protect\citeauthoryear{{Ng}}{{Ng}}{2018}]{ng.machinelearningyearning}
{Ng} A.,  2018, {Machine Learning Yearning}.
\url {https://info.deeplearning.ai/machine-learning-yearning-book#MYL-form}

\bibitem[\protect\citeauthoryear{{Papakonstantinou}, {Nikbakht}  \&
  {Eshra}}{{Papakonstantinou} et~al.}{2020}]{2020arXiv200700180P}
{Papakonstantinou} K.~G.,  {Nikbakht} H.,   {Eshra} E.,  2020, \mn@doi [arXiv
  e-prints] {10.48550/arXiv.2007.00180}, \href
  {https://ui.adsabs.harvard.edu/abs/2020arXiv200700180P} {p. arXiv:2007.00180}

\bibitem[\protect\citeauthoryear{{Paxton}, {Bildsten}, {Dotter}, {Herwig},
  {Lesaffre}  \& {Timmes}}{{Paxton} et~al.}{2011}]{Paxton2011}
{Paxton} B.,  {Bildsten} L.,  {Dotter} A.,  {Herwig} F.,  {Lesaffre} P.,
  {Timmes} F.,  2011, \mn@doi [\apjs] {10.1088/0067-0049/192/1/3}, \href
  {https://ui.adsabs.harvard.edu/abs/2011ApJS..192....3P} {192, 3}

\bibitem[\protect\citeauthoryear{{Paxton} et~al.,}{{Paxton}
  et~al.}{2013}]{Paxton2013}
{Paxton} B.,  et~al., 2013, \mn@doi [\apjs] {10.1088/0067-0049/208/1/4}, \href
  {https://ui.adsabs.harvard.edu/abs/2013ApJS..208....4P} {208, 4}

\bibitem[\protect\citeauthoryear{{Paxton} et~al.,}{{Paxton}
  et~al.}{2015}]{Paxton2015}
{Paxton} B.,  et~al., 2015, \mn@doi [\apjs] {10.1088/0067-0049/220/1/15}, \href
  {https://ui.adsabs.harvard.edu/abs/2015ApJS..220...15P} {220, 15}

\bibitem[\protect\citeauthoryear{{Paxton} et~al.,}{{Paxton}
  et~al.}{2018}]{Paxton2018}
{Paxton} B.,  et~al., 2018, \mn@doi [\apjs] {10.3847/1538-4365/aaa5a8}, \href
  {https://ui.adsabs.harvard.edu/abs/2018ApJS..234...34P} {234, 34}

\bibitem[\protect\citeauthoryear{{Paxton} et~al.,}{{Paxton}
  et~al.}{2019}]{Paxton2019}
{Paxton} B.,  et~al., 2019, \mn@doi [\apjs] {10.3847/1538-4365/ab2241}, \href
  {https://ui.adsabs.harvard.edu/abs/2019ApJS..243...10P} {243, 10}

\bibitem[\protect\citeauthoryear{Pedregosa et~al.,}{Pedregosa
  et~al.}{2011}]{scikit-learn}
Pedregosa F.,  et~al., 2011, Journal of Machine Learning Research, 12, 2825

\bibitem[\protect\citeauthoryear{{Reimers}}{{Reimers}}{1975}]{1975MSRSL...8..369R}
{Reimers} D.,  1975, Memoires of the Societe Royale des Sciences de Liege,
  \href {https://ui.adsabs.harvard.edu/abs/1975MSRSL...8..369R} {8, 369}

\bibitem[\protect\citeauthoryear{{Riess} et~al.,}{{Riess}
  et~al.}{2021a}]{2021arXiv211204510R}
{Riess} A.~G.,  et~al., 2021a, arXiv e-prints, \href
  {https://ui.adsabs.harvard.edu/abs/2021arXiv211204510R} {p. arXiv:2112.04510}

\bibitem[\protect\citeauthoryear{{Riess}, {Casertano}, {Yuan}, {Bowers},
  {Macri}, {Zinn}  \& {Scolnic}}{{Riess} et~al.}{2021b}]{2021ApJ...908L...6R}
{Riess} A.~G.,  {Casertano} S.,  {Yuan} W.,  {Bowers} J.~B.,  {Macri} L.,
  {Zinn} J.~C.,   {Scolnic} D.,  2021b, \mn@doi [\apjl]
  {10.3847/2041-8213/abdbaf}, \href
  {https://ui.adsabs.harvard.edu/abs/2021ApJ...908L...6R} {908, L6}

\bibitem[\protect\citeauthoryear{{Rocha} et~al.,}{{Rocha}
  et~al.}{2022}]{2022arXiv220316683A}
{Rocha} K.~A.,  et~al., 2022, \mn@doi [\apj] {10.3847/1538-4357/ac8b05}, \href
  {https://ui.adsabs.harvard.edu/abs/2022ApJ...938...64R} {938, 64}

\bibitem[\protect\citeauthoryear{{Sakai}}{{Sakai}}{1999}]{1999IAUS..183...48S}
{Sakai} S.,  1999, in {Sato} K.,  ed., ~ Vol. 183, Cosmological Parameters and
  the Evolution of the Universe. p.~48

\bibitem[\protect\citeauthoryear{{Salaris}, {Chieffi}  \&
  {Straniero}}{{Salaris} et~al.}{1993}]{1993ApJ...414..580S}
{Salaris} M.,  {Chieffi} A.,   {Straniero} O.,  1993, \mn@doi [\apj]
  {10.1086/173105}, \href
  {https://ui.adsabs.harvard.edu/abs/1993ApJ...414..580S} {414, 580}

\bibitem[\protect\citeauthoryear{{Saltas} \& {Tognelli}}{{Saltas} \&
  {Tognelli}}{2022}]{2022MNRAS.514.3058S}
{Saltas} I.~D.,  {Tognelli} E.,  2022, \mn@doi [\mnras]
  {10.1093/mnras/stac1546}, \href
  {https://ui.adsabs.harvard.edu/abs/2022MNRAS.514.3058S} {514, 3058}

\bibitem[\protect\citeauthoryear{Sandquist \& Bolte}{Sandquist \&
  Bolte}{2004}]{Sandquist:2004wf}
Sandquist E.~L.,  Bolte M.,  2004, \mn@doi [Astrophys. J.] {10.1086/422134},
  611, 323

\bibitem[\protect\citeauthoryear{{Schaerer}, {Meynet}, {Maeder}  \&
  {Schaller}}{{Schaerer} et~al.}{1993}]{1993A&AS...98..523S}
{Schaerer} D.,  {Meynet} G.,  {Maeder} A.,   {Schaller} G.,  1993, \aaps, \href
  {https://ui.adsabs.harvard.edu/abs/1993A&AS...98..523S} {98, 523}

\bibitem[\protect\citeauthoryear{{Serenelli}, {Weiss}, {Cassisi}, {Salaris}  \&
  {Pietrinferni}}{{Serenelli} et~al.}{2017}]{2017A&A...606A..33S}
{Serenelli} A.,  {Weiss} A.,  {Cassisi} S.,  {Salaris} M.,   {Pietrinferni} A.,
   2017, \mn@doi [\aap] {10.1051/0004-6361/201731004}, \href
  {https://ui.adsabs.harvard.edu/abs/2017A&A...606A..33S} {606, A33}

\bibitem[\protect\citeauthoryear{{Sokal}}{{Sokal}}{1997}]{Sokal}
{Sokal} A.,  1997, in {DeWitt-Morette} C.,  {Cartier} P.,   {Folacci} A.,  eds,
  NATO ASI Series, Functional Integration: Basics and Applications.
Springer, Boston, MA, USA, pp 131--192

\bibitem[\protect\citeauthoryear{Soltis, Casertano  \& Riess}{Soltis
  et~al.}{2021}]{Soltis:2020gpl}
Soltis J.,  Casertano S.,   Riess A.~G.,  2021, \mn@doi [Astrophys. J. Lett.]
  {10.3847/2041-8213/abdbad}, 908, L5

\bibitem[\protect\citeauthoryear{Stanford, Da~Costa, Norris  \&
  Cannon}{Stanford et~al.}{2006}]{Stanford2006}
Stanford L.~M.,  Da~Costa G.~S.,  Norris J.~E.,   Cannon R.~D.,  2006, \mn@doi
  [arXiv e-prints] {10.48550/arXiv.astro-ph/0605612}, \href
  {https://ui.adsabs.harvard.edu/abs/2006astro.ph..5612S} {}

\bibitem[\protect\citeauthoryear{{Streich}, {de Jong}, {Bailin}, {Goudfrooij},
  {Radburn-Smith}  \& {Vlajic}}{{Streich} et~al.}{2014}]{2014A&A...563A...5S}
{Streich} D.,  {de Jong} R.~S.,  {Bailin} J.,  {Goudfrooij} P.,
  {Radburn-Smith} D.,   {Vlajic} M.,  2014, \mn@doi [\aap]
  {10.1051/0004-6361/201220956}, \href
  {https://ui.adsabs.harvard.edu/abs/2014A&A...563A...5S} {563, A5}

\bibitem[\protect\citeauthoryear{{Sui}, {Zhang}, {Huan}  \& {Hong}}{{Sui}
  et~al.}{2019}]{2019SPIE11198E..13S}
{Sui} Y.,  {Zhang} X.,  {Huan} J.,   {Hong} H.,  2019, in {Jiang} X.,  {Chen}
  Z.,   {Chen} G.,  eds,  Society of Photo-Optical Instrumentation Engineers
  (SPIE) Conference Series Vol. 11198, Fourth International Workshop on Pattern
  Recognition. p. 1119813, \mn@doi{10.1117/12.2540457}

\bibitem[\protect\citeauthoryear{Van~Rossum \& Drake}{Van~Rossum \&
  Drake}{2009}]{python3}
Van~Rossum G.,  Drake F.~L.,  2009, Python 3 Reference Manual.
CreateSpace, Scotts Valley, CA

\bibitem[\protect\citeauthoryear{Verde, Treu  \& Riess}{Verde
  et~al.}{2019}]{Verde:2019ivm}
Verde L.,  Treu T.,   Riess A.~G.,  2019, \mn@doi [Nature Astron.]
  {10.1038/s41550-019-0902-0}, 3, 891

\bibitem[\protect\citeauthoryear{{Viaux}, {Catelan}, {Stetson}, {Raffelt},
  {Redondo}, {Valcarce}  \& {Weiss}}{{Viaux}
  et~al.}{2013}]{2013A&A...558A..12V}
{Viaux} N.,  {Catelan} M.,  {Stetson} P.~B.,  {Raffelt} G.~G.,  {Redondo} J.,
  {Valcarce} A.~A.~R.,   {Weiss} A.,  2013, \mn@doi [\aap]
  {10.1051/0004-6361/201322004}, \href
  {https://ui.adsabs.harvard.edu/abs/2013A&A...558A..12V} {558, A12}

\bibitem[\protect\citeauthoryear{{Villanova}, {Geisler}, {Gratton}  \&
  {Cassisi}}{{Villanova} et~al.}{2014}]{2014ApJ...791..107V}
{Villanova} S.,  {Geisler} D.,  {Gratton} R.~G.,   {Cassisi} S.,  2014, \mn@doi
  [\apj] {10.1088/0004-637X/791/2/107}, \href
  {https://ui.adsabs.harvard.edu/abs/2014ApJ...791..107V} {791, 107}

\bibitem[\protect\citeauthoryear{{W}es {M}c{K}inney}{{W}es
  {M}c{K}inney}{2010}]{pandasDataStructure}
{W}es {M}c{K}inney 2010, in {S}t\'efan van~der {W}alt {J}arrod {M}illman eds,
  {P}roceedings of the 9th {P}ython in {S}cience {C}onference. pp 56 -- 61,
  \mn@doi{10.25080/Majora-92bf1922-00a}

\bibitem[\protect\citeauthoryear{{Worthey} \& {Lee}}{{Worthey} \&
  {Lee}}{2011}]{2011ApJS..193....1W}
{Worthey} G.,  {Lee} H.-c.,  2011, \mn@doi [\apjs] {10.1088/0067-0049/193/1/1},
  \href {https://ui.adsabs.harvard.edu/abs/2011ApJS..193....1W} {193, 1}

\bibitem[\protect\citeauthoryear{{Yamada}, {Ohno}  \& {Ohta}}{{Yamada}
  et~al.}{2022}]{2022EP&S...74...86Y}
{Yamada} T.,  {Ohno} K.,   {Ohta} Y.,  2022, \mn@doi [Earth, Planets and Space]
  {10.1186/s40623-022-01645-y}, \href
  {https://ui.adsabs.harvard.edu/abs/2022EP&S...74...86Y} {74, 86}

\bibitem[\protect\citeauthoryear{{Yuan}, {Riess}, {Macri}, {Casertano}  \&
  {Scolnic}}{{Yuan} et~al.}{2019}]{2019ApJ...886...61Y}
{Yuan} W.,  {Riess} A.~G.,  {Macri} L.~M.,  {Casertano} S.,   {Scolnic} D.~M.,
  2019, \mn@doi [\apj] {10.3847/1538-4357/ab4bc9}, \href
  {https://ui.adsabs.harvard.edu/abs/2019ApJ...886...61Y} {886, 61}

\bibitem[\protect\citeauthoryear{pandas~development team}{pandas~development
  team}{2020}]{pandasSoftware}
pandas~development team T.,  2020, pandas-dev/pandas: Pandas,
  \mn@doi{10.5281/zenodo.3509134}, \url
  {https://doi.org/10.5281/zenodo.3509134}

\makeatother
\end{thebibliography}
\bibliographystyle{mnras}
\end{document}